\documentclass{jfm}

\usepackage{graphicx}
\usepackage{natbib}
\usepackage{bm}
\usepackage{amsmath}

\ifCUPmtlplainloaded \else
  \checkfont{eurm10}
  \iffontfound
    \IfFileExists{upmath.sty}
      {\typeout{^^JFound AMS Euler Roman fonts on the system,
                   using the 'upmath' package.^^J}%
       \usepackage{upmath}}
      {\typeout{^^JFound AMS Euler Roman fonts on the system, but you
                   dont seem to have the}%
       \typeout{'upmath' package installed. JFM.cls can take advantage
                 of these fonts,^^Jif you use 'upmath' package.^^J}%
      }
  \else
  \fi
\fi

\ifCUPmtlplainloaded \else
  \checkfont{msam10}
  \iffontfound
    \IfFileExists{amssymb.sty}
      {\typeout{^^JFound AMS Symbol fonts on the system, using the
                'amssymb' package.^^J}%
       \usepackage{amssymb}%
         \let\leq=\leqslant
         
      }{}
  \fi
\fi

\ifCUPmtlplainloaded \else
  \IfFileExists{amsbsy.sty}
    {\typeout{^^JFound the 'amsbsy' package on the system, using it.^^J}%
     \usepackage{amsbsy}}
    {}
\fi

\newsavebox{\astrutbox}
\sbox{\astrutbox}{\rule[-5pt]{0pt}{20pt}}

\newcommand{\lp}{\left(}
\newcommand{\rp}{\right)}

\newcommand{\be}{\begin{equation}}
\newcommand{\ee}{\end{equation}}
\newcommand{\bea}{\begin{eqnarray}}
\newcommand{\eea}{\end{eqnarray}}

\title[Microbubbly drag reduction in Taylor-Couette flow \dots]{Microbubbly drag reduction in Taylor-Couette flow in the wavy vortex regime}

\author[K. SUGIYAMA, E. CALZAVARINI and D. LOHSE]
{KAZUYASU SUGIYAMA$^1$\footnote{k.sugiyama@tnw.utwente.nl}, 
ENRICO CALZAVARINI$^1$\footnote{e.calzavarini@tnw.utwente.nl}
 and DETLEF LOHSE$^1$\footnote{d.lohse@tnw.utwente.nl}}

\affiliation{$^1$
Physics of Fluids group, Department of Applied Physics, J. M. Burgers Centre for Fluid Dynamics, and Impact-, MESA-, and BMTI-Institutes, University of Twente, P. O. Box 217, 7500 AE Enschede, The Netherlands.}

\date{\today}
\begin{document}
\maketitle

\begin{abstract}
We investigate the effect of microbubbles on Taylor-Couette flow by means of direct numerical simulations. 
We employ an Eulerian-Lagrangian approach with a gas-fluid coupling based on the point-force approximation. Added mass, drag, lift, and gravity are taken into account in the modeling of the motion of the individual bubble.
We find that very dilute suspensions of small non-deformable bubbles (volume void fraction below 1$\%$, zero Weber number and bubble Reynolds number $\lesssim 10$) induce a robust statistically steady drag reduction (up to 20$\%$) in the  wavy vortex flow regime ($Re=600 \textrm{ - } 2500$). 
The Reynolds number dependence of the normalized torque (the so-called Torque Reduction Ratio (TRR) which corresponds to the drag reduction) is consistent with a recent series of experimental measurements performed
by Murai \textit{et al.} (J. Phys. Conf. Ser. {\bf 14}, 143 (2005)).
Our analysis suggests that the physical mechanism for the torque reduction in this regime
is due to the local axial forcing, induced by rising bubbles, that is able to break the highly dissipative Taylor wavy vortices in the system.
We finally show that the lift force acting on the bubble is crucial in this process. When neglecting it, the bubbles preferentially accumulate near the inner cylinder and the bulk flow is less efficiently modified.
\end{abstract}

\section{Introduction}
Drag reduction induced by injection of small concentration of gas bubbles in a liquid flow has been addressed in several physical systems, from turbulent boundary layer on a flat plate (\cite{madavan84}, \cite{fer04}, \cite{sanders06}), to  channel flows (\cite{xu02},  \cite{lu05}). 
Despite the efforts dedicated to the subject, the problem has not been
fully clarified from a fundamental physical point of view. Different
mechanisms such as effective compressibility of the flow (\cite{fer04}),
bubble deformability (\cite{lu05}), or compressibility (\cite{lo06}), or splitting 
(\cite{meng98}, have been proposed as relevant. 
A misleading aspect of the drag-reduction problem arises from the fact that, depending on the system considered, either transient, or spatially dependent (as a function of the bubbles' injection point), or statistically steady effects can be observed. To overcome such a difficulty, more recently experiments have been conducted in a Taylor-Couette (TC) set-up, (\cite{djeridi04},  \cite{vandenberg05}, \cite{murai05}, \cite{vandenberg07}). The TC system has two advantages. 
First, since it is a closed system, exact global energy balance relation can be explicitly obtained and drag variations can be evaluated in terms of a single globally averaged quantity, the torque.
Second, statistically stationary states can be reached easily.
In order to quantify the level of drag reduction, it is convenient to consider the so called Torque Reduction Ratio (TRR) coefficient, viz.
\be \label{eq:TRR} 
\textrm{TRR} \equiv 1 - \frac{T_{b}}{T},
\ee
where $T_b$ stands for the measured torque in the two-phase system, i.e., with the bubbles included and $T$ for the torque in the single-phase flow at the same Reynolds number.
In the highly turbulent TC system ($Re > 10^{5}$) drag reduction is mainly attributed to the bubble deformation mechanism (\cite{vandenberg05}, \cite{lu05}) in the boundary layer (\cite{vandenberg07}).

Also at moderate flow-Reynolds numbers in the range $Re = 600  \textrm{ - } 4500$ (the so called wavy/modulate-wavy Taylor vortex regime)  
 \cite{murai05} reported that bubble-induced drag reduction up to $25\%$ can be observed for tiny bubble concentrations, $O(0.1\%)$ in volume. 
The experimental apparatus considered by these authors is forced by rotation of the inner cylinder, while the outer one is kept fixed. 
The cylindrical enclosure is filled by a Silicon oil roughly five times
 more viscous than water $(\nu = 5 \cdot 10^{-6}\ {\rm m}^2/{\rm s})$
 and air bubbles are injected into the fluid. Their typical radius is
 $a= 250 \textrm{ - } 300$ $\mu {\rm m}$ and Weber number $We \equiv 2 a \rho v_T^2 \sigma^{-1} < 0.6$ (where $\rho$ and $\sigma$ are respectively the density and the surface tension of the oil and $v_T$ the bubble terminal velocity in still fluid). 
The bubble Reynolds number, based on $v_T$, is $Re_b = 2 a\ v_T / \nu \simeq 7$.
Therefore bubbles can be  regarded with good approximation as mono-disperse 
non-deformable spheres, whose scale is of the same order as the smallest scales of fluid fluctuations. 
The experimental results of Murai \textit{et al.} showed that the drag reduction decreases with
increasing Reynolds number. TRR was almost vanishing at $Re \simeq 4000$ and became negative (drag increase) for even larger Reynolds numbers.  Murai \textit{et al.} explained this drag reduction effect through the bubble interaction with the coherent Taylor vortices, producing a vertical elongation of their arrangement.

The features of the Murai \textit{et al.} experiment, in particular the low large-scale $Re$ numbers and the small (i.e. limited $Re_b$) non-deformable bubbles employed there, make this experiment accessible to a comparison with numerical simulations, which we will perform in this paper.
 
The aim of the present work is to shed more light on the physical mechanism of drag reduction observed in the flow conditions of Murai \textit{et al.} experiment.
However, given the complexity of this two-phase flow system, appropriate modeling of the dynamics 
is unavoidable 
and approximations must be introduced.
First, we use an Eulerian-Lagrangian algorithm.
The forces on each point-bubble are modeled through
 effective drag and lift forces; they act back on the fluid through
two-way coupling.
Second, in order to make the simulation computationally feasible, we have to restrict the flow to a smaller domain as compared to the experiment and to adopt different, i.e.\ periodic, boundary conditions. 
Third, in order to achieve good convergence, the bubbles are initially
positioned at random positions throughout the flow. 
This is obviously different from the experimental procedure.
In the experiment bubbles are injected from the bottom. 
The injection sites are close to the inner cylinder walls.  The bubbles emerge to the free surface at the top after traveling across the cylindrical enclosure, whose aspect ratio is much larger than in the simulation (cf.\ the $\Gamma$  values in Table\ \ref{tab1}). 
It is finally on the top TC section that the radial mean void fraction is measured. 
 The experimenters   assumed  that the bubble distribution rapidly reaches a fully developed state that does not keep memory of the inhomogeneous injection sites.
This assumption justifies the different bubble injection procedure adopted in our numerics.

In spite of these unavoidable differences, we think -- and we will further
motivate in the paper -- that
the main flow features and the resulting drag reduction 
phenomena are reasonably captured by our numerical approach. 
In particular, it is of relevance to note that drag reduction mechanism observed in the flow conditions examined here -- which we will call \textit{pseudo-turbulent} drag reduction -- turns out to be substantially different from the mechanism 
that  acts under the highly-turbulent conditions explored by \cite{vandenberg05,vandenberg07}. Under those conditions the bubble deformation had turned out to 
be crucial,  see also \cite{lu05}.

The paper is organized as follows: in section \ref{sec:system}, we introduce the geometry and the two-phase model system used for the numerical study and present a test in order to validate the code.
In section \ref{sec:results} our numerical results on torque modulation in the two-phase flow and bubble distributions are described and discussed in detail. Section \ref{sec:conclusions} contains comments and conclusions. In the appendix we derive relations connecting globally averaged quantities in the TC system. 
\begin{figure}
\begin{center}
\includegraphics[width=.35\textwidth,angle=0]{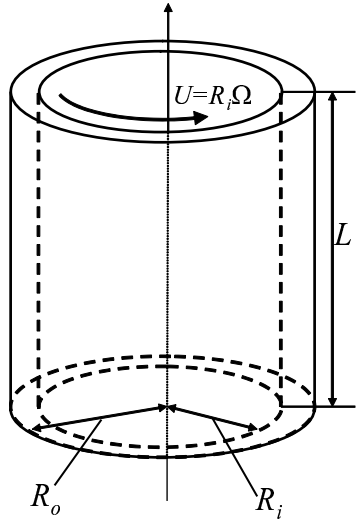}
\caption{Sketch of the Taylor-Couette system}\label{fig:setup}
\end{center}
\end{figure}

 \section{Taylor-Couette system and numerical scheme}\label{sec:system}
\subsection{Definitions}
In the Taylor-Couette system the fluid is enclosed between two coaxially rotating cylinders  with radius  $R_o$ and $R_i$, respectively for the outer and the inner one. The height of the cylindrical enclosures, denoted with $L$,  is in the vertical direction. We limit our investigation to the case where only the inner cylinder is rotating at a constant angular velocity $\Omega$ while the outer one is at rest, see Fig. \ref{fig:setup}. 
Furthermore, no-slip conditions are assumed at the walls.
It is convenient to adopt the Reynolds number based on the inner cylinder velocity $U=R_i \Omega$ and the gap width $d=(R_o-R_i)$ as a control parameter, 
that is
\be\label{eq:Re}
Re \equiv \frac{(R_o - R_i)R_i\Omega}{\nu}.
\ee
The 
response of the system is the time averaged torque on the inner cylinder, which is expressed as
\be\label{eq:T}
T = 2 \pi R_i^2 L \tau_{wi},
\ee
where $\tau_{wi}$ stands for the mean inner-wall shear stress. 
In cylindrical coordinates $(z,r,\theta)$ the wall shear stress 
can be computed as 
\be\label{eq:tw}
\tau_{wi} = - \rho \nu \left( \frac{{\rm d}}{{\rm d}r} \overline{u_{\theta}}  |_{r=R_i}  - \Omega  \right),
\ee
where $u_{\theta}$ is the azimuthal velocity component and the over-bar denotes averaging in space over concentric cylindrical surfaces and time. Note that an analogous expression for the torque can be written in terms of the outer
cylinder, since conservation of angular momentum implies
 $( \frac{{\rm d}}{{\rm d}r} \overline{u_{\theta} }  |_{r=R_o} ) R_o^2 =
 ( \frac{{\rm d}}{{\rm d}r} \overline{u_{\theta}} |_{r=R_i} - \Omega ) R_i^2$.
At increasing angular velocity the flow in between the two cylinders experiences several hydrodynamic instabilities. After the destabilization of the laminar flow, the formation of horizontal couples of circular rolls, known as Taylor vortices (\cite{taylor23}), is observed. 
Then, at further increasing $\Omega$, axial
 symmetry is lost and one encounters first steady wavy, then time modulated rolls, and finally a fully developed turbulent regime  is established, see \cite{andereck86}.

The relation linking the torque to the Reynolds number and the gap parameter $\eta = R_i/R_o$
cannot be directly computed from the equation of motion, except for the laminar (Couette) regime.
To deduce under general conditions the functional relation $T(Re,\eta)$ is still an open problem, see  \cite{lathrop92},  \cite{esser96}, \cite{eckhardt00}, \cite{lim04}, and \cite{gross07}. 
However, the torque can be related to the mean (time and volume averaged $\langle \ldots \rangle$) energy dissipation rate $\langle \varepsilon \rangle$, namely 
\footnote{Note that in \cite{lathrop92} and in \cite{vandenberg05} this relation contains a typo, namely, a wrong pre-factor 2. See also the appendix.}
\be\label{eq:eps}
\langle \varepsilon \rangle = \frac{T \Omega}{\pi \rho (R_o^2 - R_i^2) L}.
\ee
Thus $\langle \varepsilon \rangle$ and $T$ are directly proportional and a decrease in the torque corresponds to a 
reduction of the internal energy dissipation rate in the system. Therefore, the total drag reduction can be deduced from the variations of a single global observable as the torque.

\begin{figure}
\begin{center}
\includegraphics[width=.43\textwidth,angle=-90]{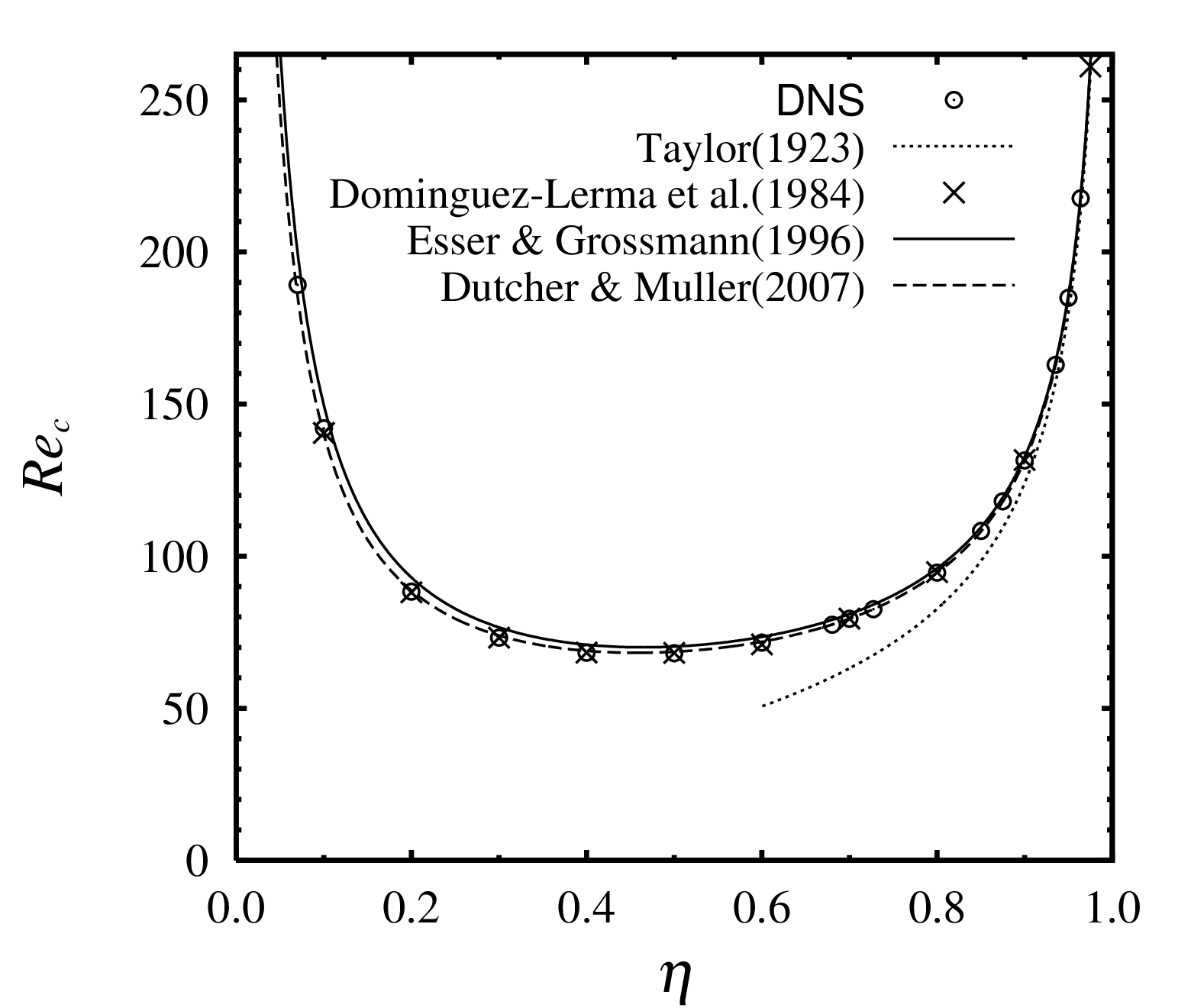}
\includegraphics[width=.43\textwidth,angle=-90]{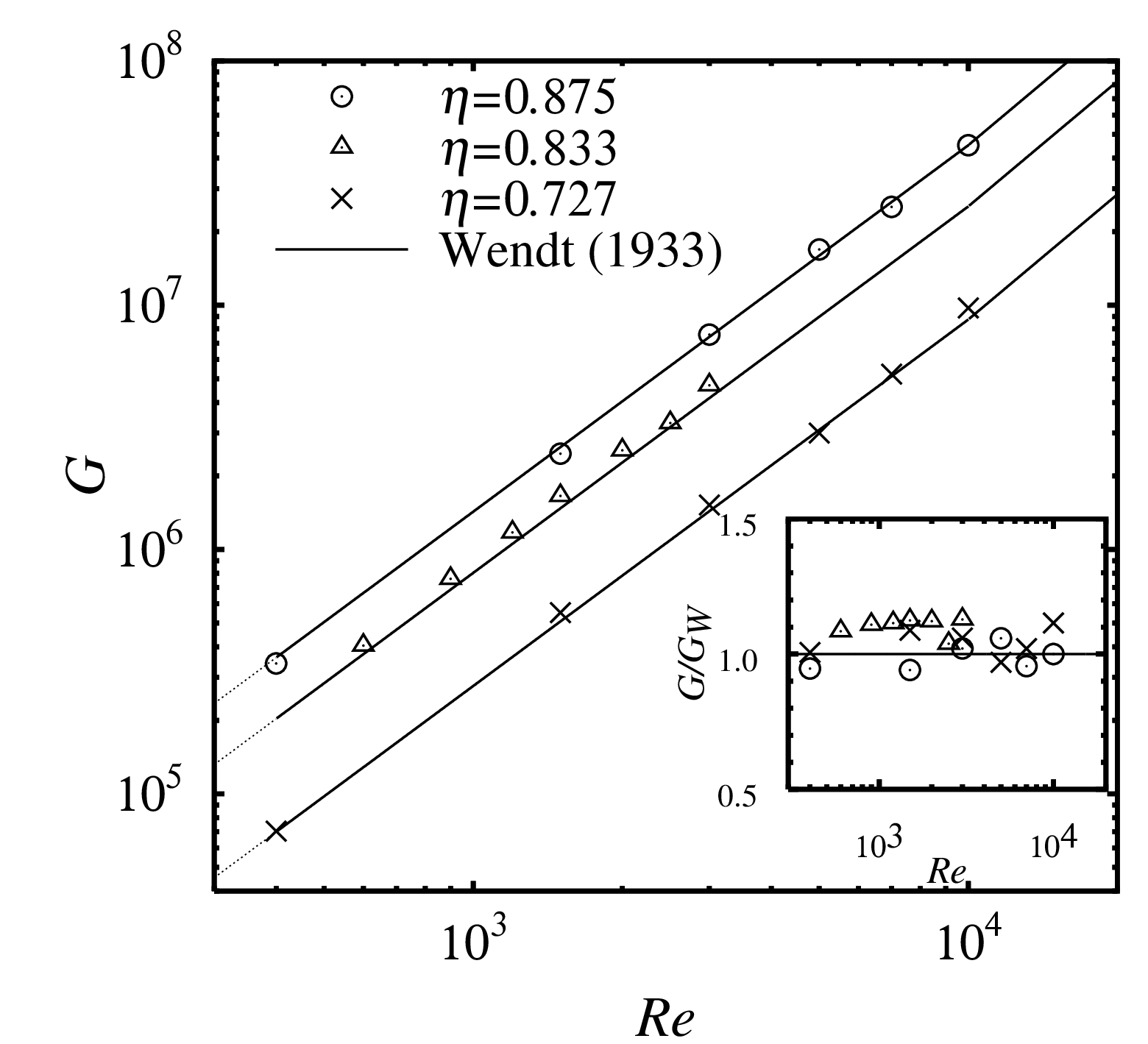}
\caption{ Single-phase flow code validation. 
(left) The critical Reynolds number $Re_c$ as a function of the gap ratio $\eta$, and comparison with 
\cite{taylor23} narrow-gap limit ($\eta \to 1$), \cite{dom84}, \cite{esser96} and \cite{dutch07} predictions.
(right) Dimensionless torque $G=T/(\rho \nu^2 L)$ versus $Re$ for three different values of the gap ratio $\eta$, and comparison with Wendt's empirical relation $G_W=1.45 ( Re\  \eta)^{3/2}/(1-\eta)^{7/4}$, \cite{wendt33} . 
The inset shows a compensated plot $G/G_W$ vs. $Re$.}\label{fig:G}
\end{center}
\end{figure}

\subsection{The model for the two-phase flow}\label{sub:model}
The direct numerical simulation (DNS) approach we adopt is based on the Eulerian-Lagrangian algorithm, here in the form used by 
\cite{mazzitelli03, mazzitelli03b}.
The fluid phase is described by the Navier-Stokes equation for an incompressible flow $( \nabla \cdot {\bf u} = 0 )$. Bubbles counter-act on the flow by a \textit{two-way} point-like coupling:  
\be
\label{eq:ns}
\partial_t {\bf u} + \left(  {\bf u} \cdot \nabla \right) {\bf u} = - {{\nabla p} \over {\rho}} +\nu\ \nabla^2 {\bf u} + {\bf f}_b,
\ee
\be \label{eq:twoway} 
{\bf f}_b = \sum_i^{N_b} \frac{4 \pi a^3}{3}  \delta({\bf x} - {\bf y_{(i)}}) 
\lp  \partial_t {\bf u} + \left(  {\bf u} \cdot \nabla \right) {\bf u} - {\bf g}  \rp,
\ee
where  ${\bf y_{(i)}}$ denotes the instantaneous position of the $i$-th bubble and ${\bf g}$ the gravitational acceleration.
The bubbles are tracked as Lagrangian particles with a model equation for which added mass, drag, lift, and gravity are taken into account.
Their dynamics is given by \cite{maxeyriley83}, \cite{auton87}, \cite{auton88}, \cite{climent99}, \cite{mazzitelli03b}:
\bea 
\frac{{\rm d} {\bf y}}{{\rm d}t} &=& {\bf v}, \label{eq:bubble1}\\
\frac{{\rm d} {\bf v}}{{\rm d}t} &=& 3 \left( \partial_t {\bf u} + \left(  {\bf u} \cdot \nabla \right) {\bf u} \right) - \frac{1}{\tau_b} \left( {\bf v} - {\bf u} \right)- 2 {\bf g} - \left( {\bf v} - {\bf u} \right) \times {\bm \omega}, \label{eq:bubble2}
\eea
with $\tau_b \equiv a^2/(6 \nu)$ the characteristic bubble response time, and ${\bf u}$ and ${\bm \omega}$ denoting respectively the fluid velocity and vorticity vectors at the bubble position. 
Fully-elastic boundary conditions are assumed for the bubbles on the cylinder's lateral walls, while direct bubble-bubble interactions (4-way coupling) are neglected.
It is important to note that the lift force in (\ref{eq:bubble2}) is only approximate. Here the expression for the rotational flow in the limit $Re_b \to \infty$ is used, corresponding to a lift coefficient $C_L=1/2$ (\cite{auton87}).  
However, in the simulation we consider the case of $Re_b \sim O(10)$ in an unsteady flow for which the 
Auton's expression may be incorrect or may overestimate the actual intensity of the lift force (\cite{nierop07}). 
Despite the above mentioned approximations and the simplified approach of the model adopted, it has been shown by \cite{ramon06b} and \cite{calza06},  that a qualitative agreement with real systems occurs.
We will further discuss on the effect of the lift on the bubble dispersion in the TC flow in section \ref{subsec:lift} of this paper.

\subsection{Numerical implementation, code validation and tuning of the DNS}\label{sub:numerics}
The continuous phase, (\ref{eq:ns}), has been implemented in cylindrical coordinates on a finite-difference scheme.
The grid, $N_z\times N_r \times N_{\theta} = 128\times 64 \times 256$, is non uniform and refined in the near-wall regions in the radial direction, while it is uniform both in the vertical and azimuthal ones.
Careful attention has been paid to the implementation of the non-linear term by the use of an algorithm that enjoys highly-conservative properties for energy, proposed by \cite{fukagata02}.
The time marching scheme is of second order accuracy.
The bubble trajectories (\ref{eq:bubble1})-(\ref{eq:bubble2}) have also been implemented in cylindrical coordinates as in \cite{djeridi04} and integrated by a second order scheme in time, while interpolation of fluid velocity at the bubble position is implemented by a third-order scheme, i.e. taking into account up to the second nearest lattice nodes (64 points). 
The feed-back (\ref{eq:twoway}) is extrapolated from the bubble position to the next nearest nodes by a tri-linear scheme (8 points).
We use periodic boundary conditions in the vertical direction both for the fluid and the bubbles. Additionally, for the fluid only, we fix the mean flow in the vertical direction to be zero, in order to avoid the emerging of a mean axial current that would prevent a realistic comparison with the experiment, where it is forbidden for geometrical reasons.

The implementation of the single-phase flow dynamics  has been validated by measuring the onset of the primary instability, i.e., the critical Reynolds number as a function of the gap ratio, $Re_c(\eta)$, as well as the dimensionless torque $G=T/(\rho \nu^2 L)$ at different Reynolds numbers and for different values of (the gap ratio) $\eta$. 
The results on $Re_c(\eta)$ were compared with the available analytical calculations,\cite{taylor23} (for the narrow-gap limit $\eta \to 1$), \cite{dom84} and with phenomenological estimates given by \cite{esser96} and \cite{dutch07}. 
Moreover we compare $G(\eta, Re)$ with  the empirical law given by \cite{wendt33} (also in \cite{lathrop92}), finding 
 satisfactory agreement (figure \ref{fig:G}).

The tuning of our DNS with the experimental setup adopted by Murai \textit{et al.} is then done as follows.
The experimental apparatus is geometrically characterized by a gap ratio $\eta= 5/6$ and an aspect ratio $\Gamma= L/d = 20$, see figure 1 of \cite{murai05} for a sketch of their setup.
The flow boundaries adopted in the experiment on the horizontal surfaces are free fluid surface (free-slip) at the top and zero mass-flux at the bottom.
Since we do not expect drag to be strongly affected by the influence of the top/bottom boundary conditions, 
particularly for aspect ratio values ($\Gamma$) much larger than one, we use, for simplicity and in order to limit the computational costs a smaller aspect ratio $\Gamma/d = 8$ than in experiment, and periodic boundary conditions as mentioned before. The gap ratio is as in the experiment,  $\eta=5/6$. {It turned out that the constraint of periodic boundaries to the flow may introduce a bias
on the formation and breaking of coherent structures in the flow, on this point we will comment more in detail in section 3.2.}
Furthermore, for the whole set of numerical simulation we fix the ratio $a/d$ and the bubble Reynolds number $Re_b$ to the experimental values - both these quantities are not Reynolds dependent in the experiment - and we vary the fluid Reynolds number $Re$, by changing the inner cylinder rotational speed $\Omega$ as in the experimental setup,  in the range $Re=600 \textit{-}2500$. 
Care has to be put on the tuning of the void fraction, since in the
experiment it was not kept constant among the different runs but
estimated \textit{a posteriori} through a measure of the total gas flow
rate. We chose the total void fraction $\alpha_0$ consistent with the experiment. The values
(Table\  \ref{tab1}) are obtained by a fit to the estimated experimental void fraction (see also \cite{oiwa05}, chapter 5); this explains some tiny deviations among the two cases. For the flow conditions that we consider for the comparison, i.e. $Re \leq 2500$, $\alpha_0$ is always less than $1\%$.
The value of the ratio $a/d$ and the total void fraction value $\alpha_0$, then fix the number of bubble we must use in the numerical study, namely $N_b = \alpha_0  (  \pi L (R_o^2 - R_i^2) )/ \frac{4}{3} \pi a^3$. The number of bubbles $N_b$ varies between few thousands to a maximum of roughly $1.5 \cdot10^4$.
Table \ref{tab1} summarizes the set of parameters adopted for the simulations and the experiment.
\begin{table}
\begin{center}
\begin{tabular}{c|c|c|c|c|c|l|c|c}
\hline
	       &   $\eta$       &  $\Gamma$ & $a/d$ &  b.c. & $Re$ & $\alpha_0\ (\%)$ &  $Re_b$ & $ We $\\
\hline
Exp.        &      5/6         &   20             &   1/40  &  vert. no-slip,     &   $600$ - $4500$     &  0.092 at $Re=600$         &   7.1           &   $\leq 0.6$    \\
               &                    &                    &            &  horiz. top free-slip,  &       &  0.721 at $Re=2700$        &                  &                \\
               &                    &              &            &  horiz. bottom zero-mass flux &  &  2.446 at $Re=4200$           &            &           \\
\hline
Sim.        &      5/6         &       8      &     1/40       &  vert. no-slip,  &  $600$ - $2500$   & 0.125 at $Re=600$         &  7.1          &     0      \\
               &                    &              &            & horiz. periodic      &      &    0.670 at $Re=2500$        &          &           \\
\hline
\end{tabular}
\end{center}
\caption{Relevant scales and parameters for the two-phase Taylor-Couette system studied experimentally by 
\cite{murai05} and numerically in this paper. We report first the three geometrical parameters: gap ratio $\eta$, aspect ratio $\Gamma$, and the ratio between the bubble radius and the gap width $a/d$, then the flow boundary condition, and the large scale flow Reynolds number range. Finally, bubble characteristics are listed: the global volume void fraction $\alpha_0$ (in per cent), and the bubble's typical Reynolds and Weber numbers.}\label{tab1}
\end{table}

A typical simulation run at a given value of the Reynolds number is composed of two parts. First, a single-phase DNS is performed until statistically stationarity is reached. Then bubbles are positioned in the flow with random homogeneous distribution and velocities equal to the local fluid velocity. 
This bubble injection procedure is supposed to mimic 
the presumably fully developed state
at the top TC section where the bubble distributions have been
measured in experiment. 
Simultaneously with the bubble injection,
the bubbles-fluid (\textit{two-way}) coupling is activated. 
The two-phase simulation is then performed until a statistically steady state is well established.
All relevant physical observables in the single- and two-phase flow are computed by suitable time averages, of order O($10^2$) revolution times 
($=2 \pi \Omega_i^{-1}$), in the statistically stationary regime.

\begin{figure}
\begin{center}
\includegraphics[width=.4\textwidth,angle=-90]{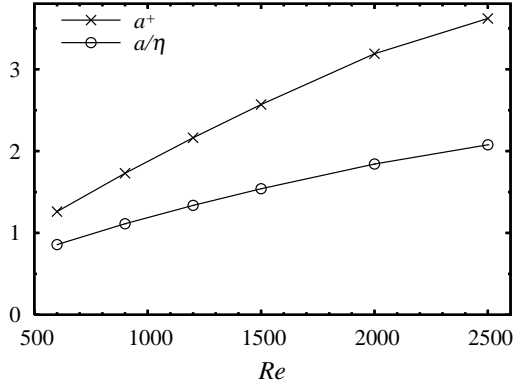}
\caption{ 
Ratio $a/\eta$
between bubble radius and Kolmogorov scale (circles) 
and 
ratio $a^{+} = a \sqrt{\tau_w / (\rho \nu^2)}$
between bubble radius and wall length unit,
both 
as a function of $Re$.
Both the Kolmogorov length $\eta$ and the wall-shear-stress $\tau_w$ are evaluated on the single-phase flows.
The thickness of the viscous-sub layer ($l_v$) in a planar channel-flow in wall-units is roughly $l_v^+ \simeq 5$. 
}\label{fig:kolm}
\end{center}
\end{figure}

Finally, in order to verify the consistency of the model adopted in the DNS, we check if the smallest scale in the system is of the same order of the bubble typical size.
In particular we compare the bubble radius $a$ with the size of the Kolmogorov dissipative scale $\eta= ( \nu^3 / \langle \varepsilon \rangle)^{1/4}$ and, alternatively, we look at the magnitude of the wall-units radial size $a^{+} \equiv a \sqrt{\tau_w / (\rho \nu^2)}$. 
While the ratio $a/\eta$ is relevant in the bulk flow the $a^{+}$ is of importance in near-wall regions.
The results reported in figure \ref{fig:kolm} show that our assumption
of point bubbles indeed is 
reasonable.

\section{Numerical results}\label{sec:results}
\subsection{Overview}\label{sub:overview}
\begin{figure}
\begin{center}
\includegraphics[width=.6\textwidth,angle=-90]{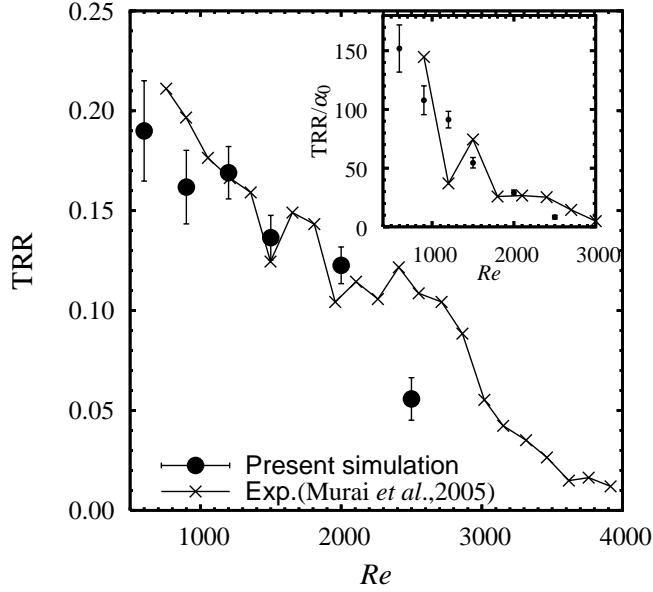}
\caption{Torque reduction ratio computed in the numerical simulations and the experimental measurements performed by \cite{murai05}. In the inset TRR is divided by the mean void fraction $\alpha_0$. Note that the void fraction is not kept constant with $Re$, see table \ref{tab1}.}\label{fig:TRR_re}
\end{center}
\end{figure}

In this section we present the comparison of numerical and experimental results.  
In figure \ref{fig:TRR_re} we show both the TRR coefficient obtained
from the experiment and from the simulations. In the inset TRR is
normalized by the total void fraction (this ratio is called sensitivity 
by \cite{murai05}). 
\begin{figure}
\begin{center}
\includegraphics[width=.45\textwidth,angle=-90]{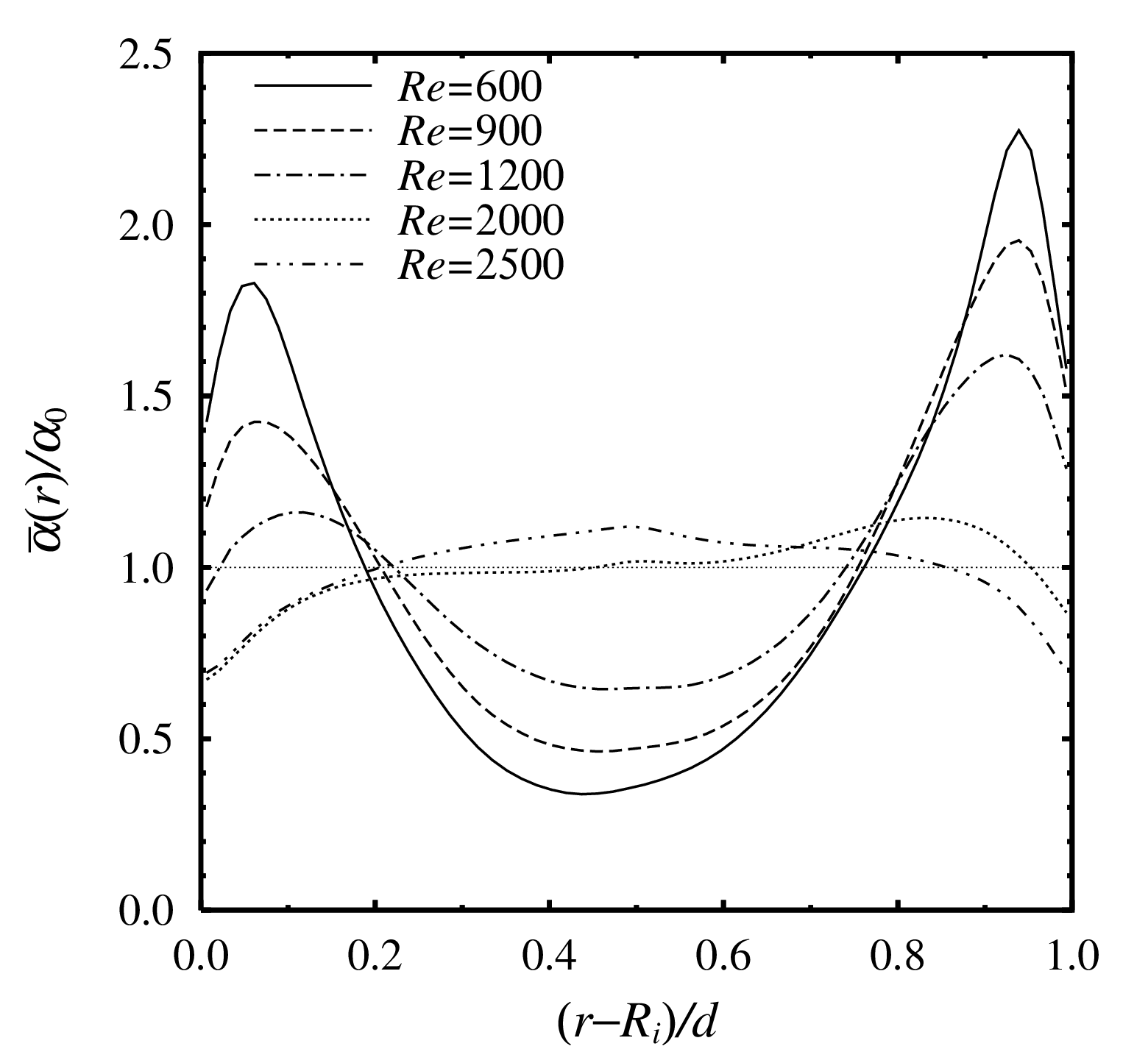}
\includegraphics[width=.45\textwidth,angle=-90]{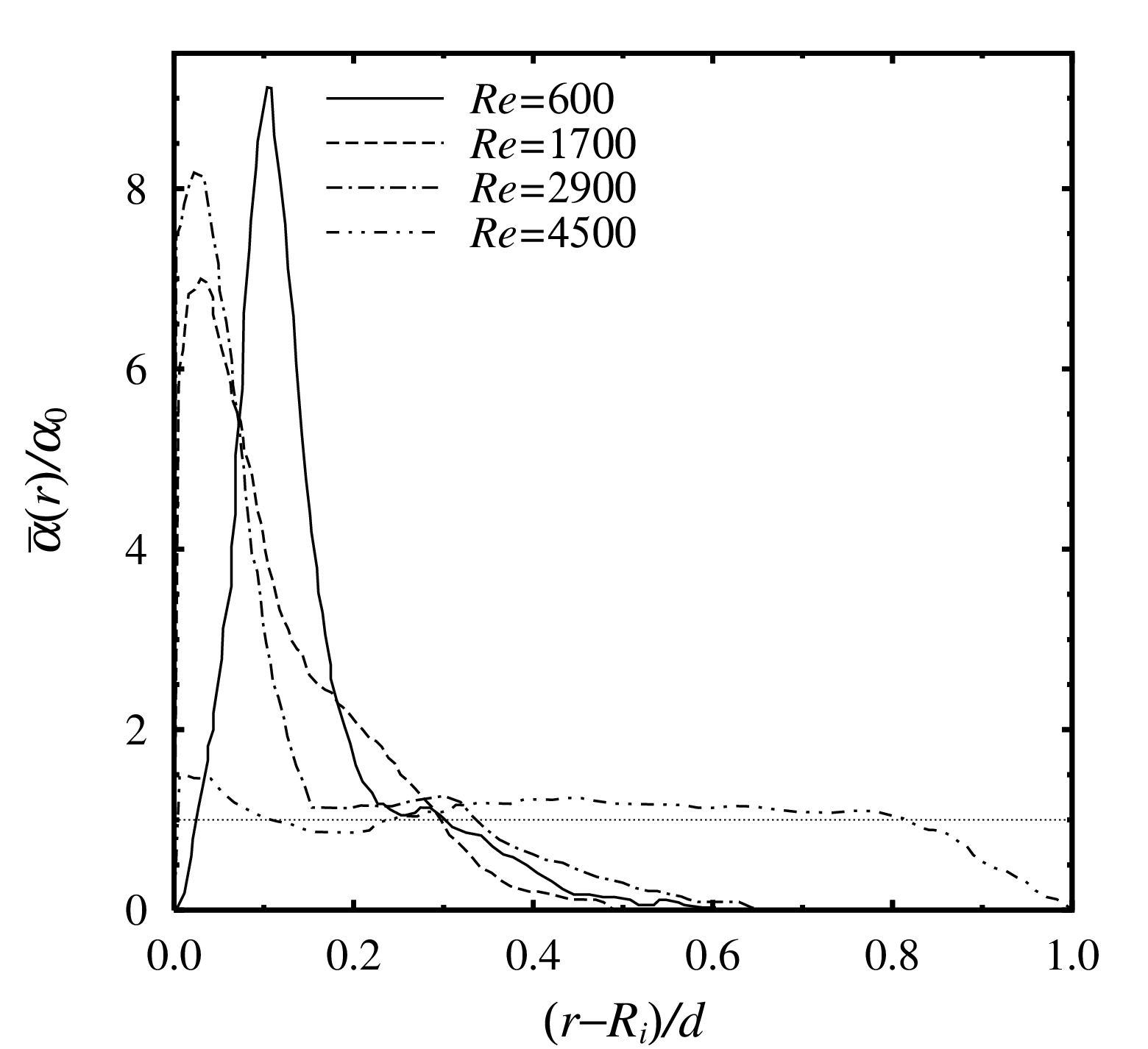}
\caption{Normalized mean local void fraction in the radial direction, $\overline{\alpha}(r)/\alpha_0$, at different $Re$ values, in the numerics (left) and the experiment (right). The quantity $\overline{\alpha}(r)/\alpha_0$ has been adopted from the data in figure 9 of \cite{murai05}. 
For better comparison the homogeneous distribution line $\overline{\alpha}(r)/\alpha_0 = 1$ is also shown.}\label{fig:void}
\end{center}
\end{figure}
The quantitative agreement on this observable is satisfactory. 
First, we detect a robust (up to 20$\%$) statistically steady  drag reduction. 
Second, drag reduction effect has a decreasing trend at increasing the Reynolds number. 
How can the two above mentioned observations be explained? Paragraphs \ref{sec:origin} and \ref{sec:decrease} of this section will be devoted to understand the physical mechanism leading to this behavior.

Less satisfactory is the comparison with the experiment of the local mean void fraction (or mean bubble concentration). In figure \ref{fig:void} we show the mean radial bubble concentration profiles normalized by the total void fraction, both from the numerical simulations and from the experiment. 
This quantity is defined as,
\be \label{eq:ar}
\frac{ \overline{\alpha}(r) }{\alpha_0}  = 
\frac{ R_o^2-R_i^2 }{ 2 N_b }\  \frac{1}{T_s} 
\int_{t_0}^{t_0 + T_s}\!\!\!\!\!\!{\rm d}t\  \sum_i^{N_b} 
\frac{ \delta( r - y_{r (i)}(t)) }{r},
\ee
where $y_{r (i)}(t)$ is the radial component of the position vector of the $i$-th bubble and $T_s$ indicates the duration of the time interval where the two-phase flow is in statistically steady conditions.
The local bubble concentration in the simulations is less inhomogeneous than in the experiment.
The maximum deviation from equi-distribution is about $100 \%$ of the homogeneous case and is peaked both at the inner and the outer wall.
On the other hand, in the experiment, where radial mean void fraction has been evaluated by averaging over a series of snapshots taken in the upper part of the TC system, a much stronger accumulation peaked near the inner cylinder is found and it reaches much larger values (roughly 4 times larger than in numerics in the most extreme case $Re=600$).
Nevertheless, the same trend towards homogenization at increasing Reynolds number is observed in both cases. 
The possible origin of this discrepancy and its relevance on the effect of drag reduction will be further discussed in detail and will be the focus of section \ref{subsec:lift}.
\begin{figure}
\begin{center}
\includegraphics[width=.35\textwidth,angle=-90]{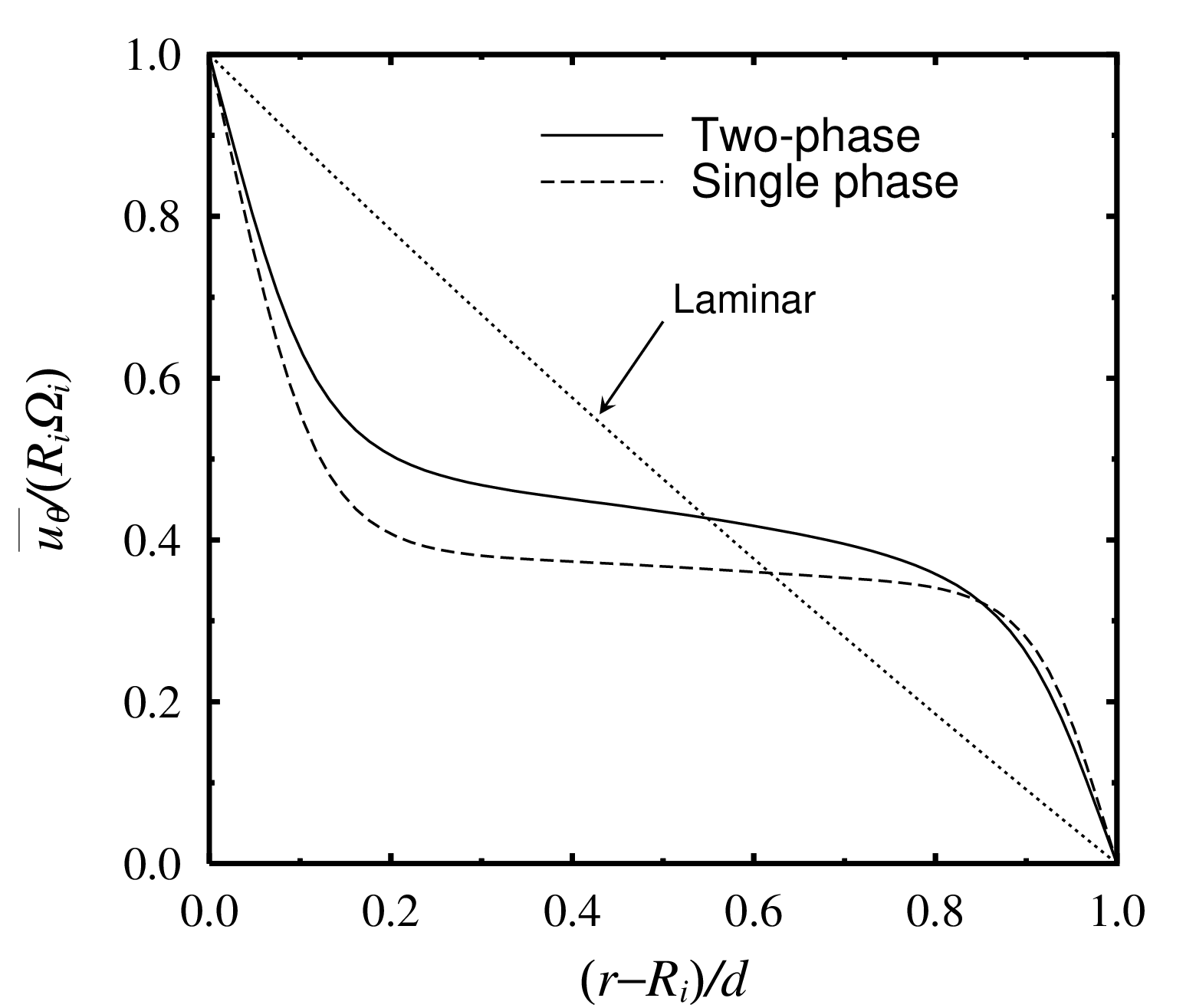}
\includegraphics[width=.35\textwidth,angle=-90]{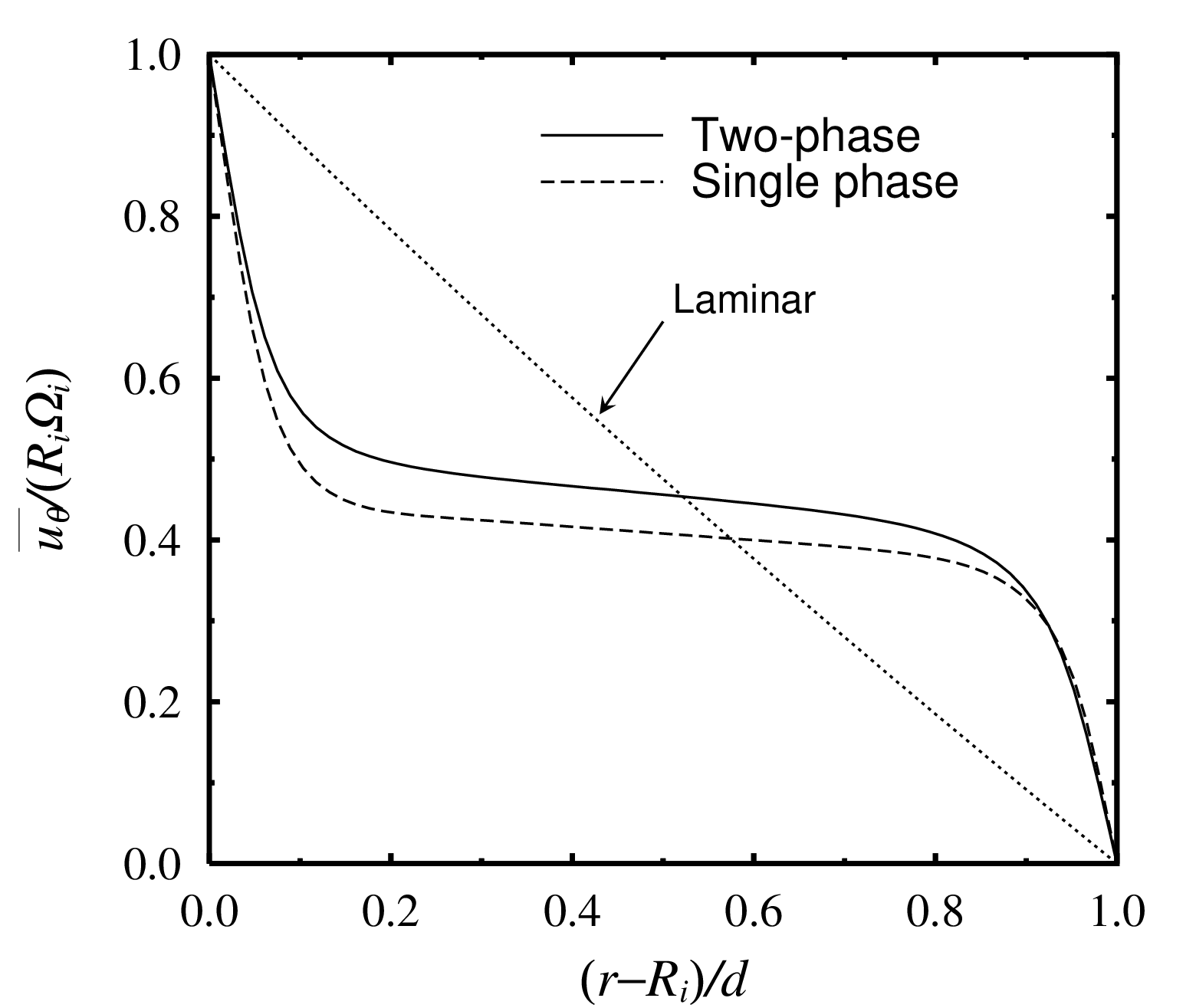}\\
\includegraphics[width=.35\textwidth,angle=-90]{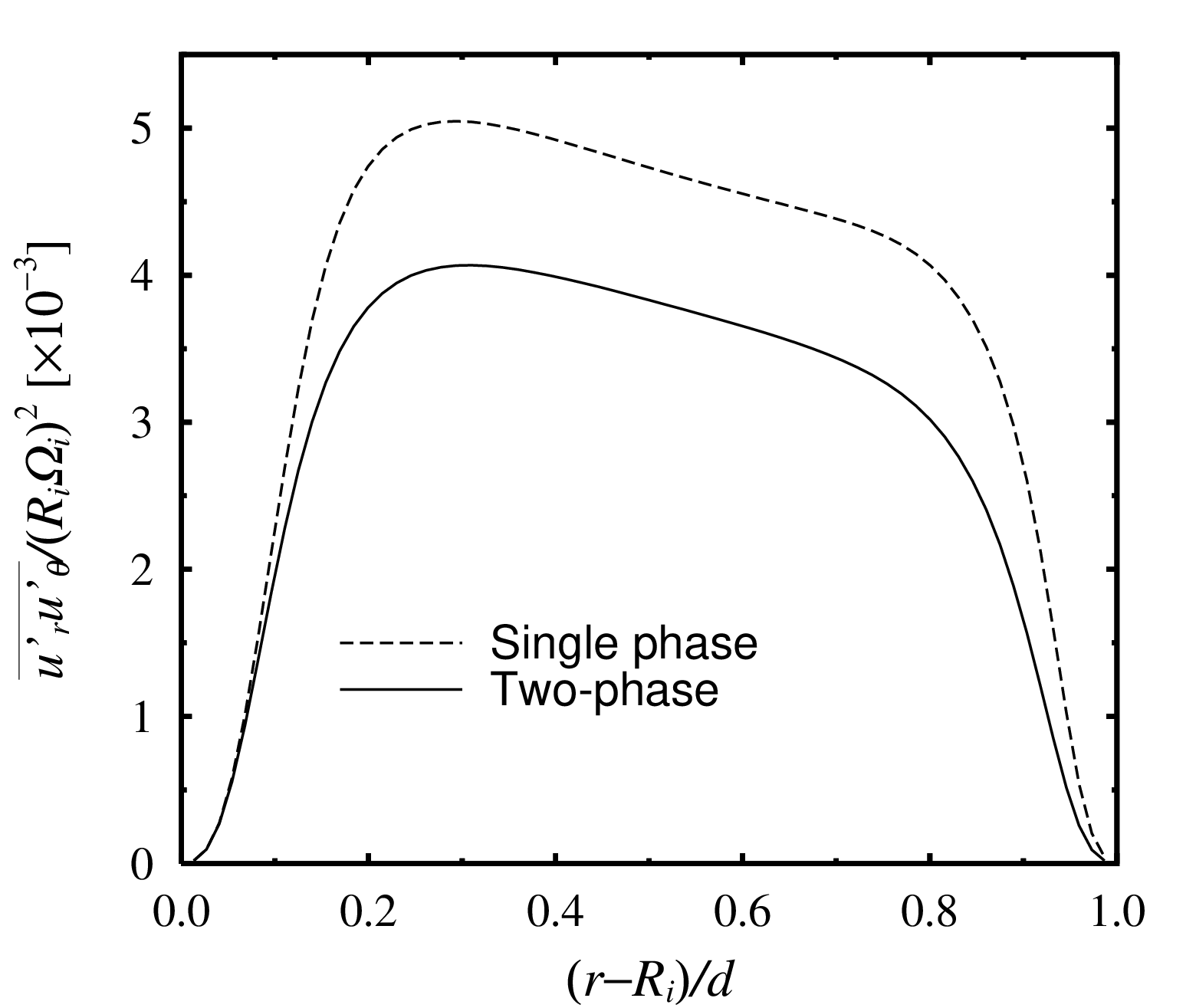}
\includegraphics[width=.35\textwidth,angle=-90]{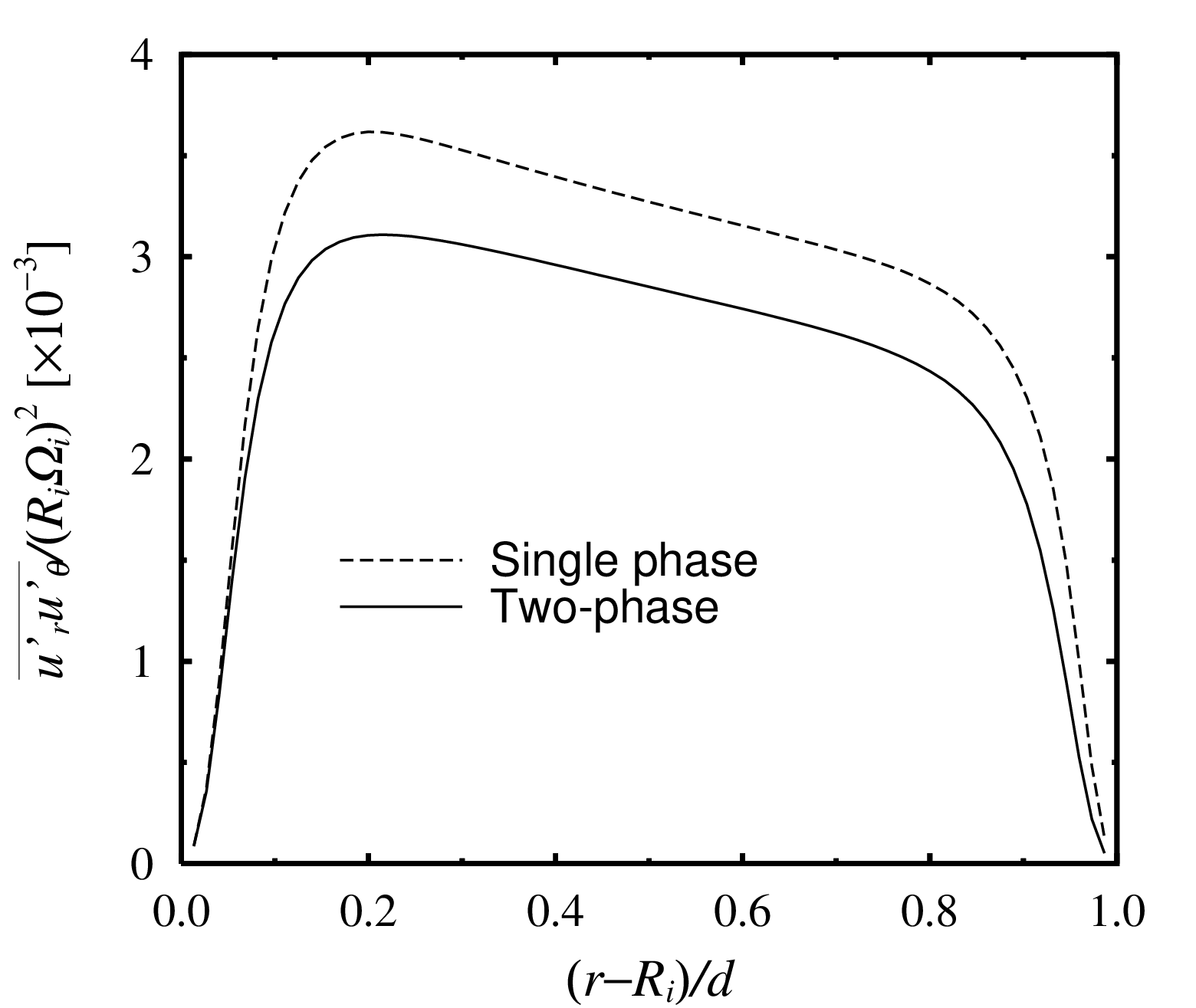}
\caption{Top panel: mean azimuthal velocity profile $\overline{u_{\theta}}(r)$ at $Re=900$ (left) and $Re=2000$ (right). The laminar profile is also reported. Bottom panel: the mean Reynolds shear stress component $\overline{u_{\theta} u_{r} }(r)$ at $Re=900$ (left) and $Re=2000$ (right).  
}\label{fig:vprofiles}
\end{center}
\end{figure}

\subsection{Origin of the drag reduction} \label{sec:origin}
\begin{figure}
\begin{center}
\includegraphics[width=.45\textwidth,angle=-90]{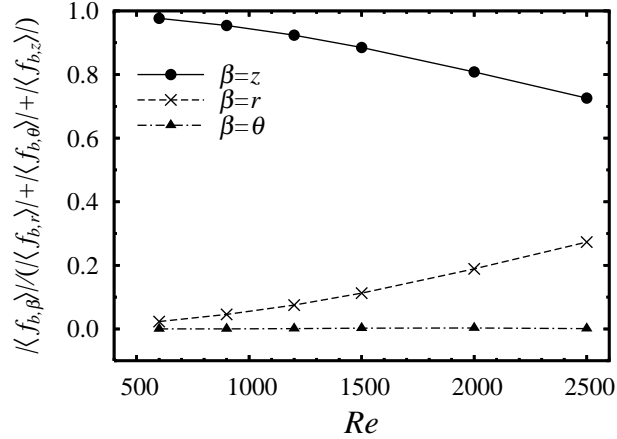}
\caption{Normalized globally averaged (space and time) relative components of the bubble forcing  $| \langle f_{b, \beta} \rangle | / ( | \langle f_{b, \theta } \rangle | +  | \langle f_{b, r } \rangle | + | \langle f_{b, z} \rangle | )$ with $\beta= z, r, \theta$ vs. $Re$.}\label{fig:fb}
\end{center}
\end{figure}

The variation of the intensity of the torque in the TC system is directly connected to a change in the mean azimuthal velocity profile along the radial direction $\overline{u_{\theta}}(r)$, see (\ref{eq:T})-(\ref{eq:tw}). Figure \ref{fig:vprofiles} (top) shows the actual modification of the two-phase profile compared to the single-phase at two different Reynolds numbers, $Re=900$ and $Re=2000$. These $Re$ values will be adopted as reference cases in the following discussion. A smoother slope is observed in figure \ref{fig:vprofiles} for the gradient at the wall of $\overline{u_{\theta}}(r)$ in the two-phase  case. As consequence the torque  is reduced. We also note that the torque $T_b$ for the two phase system, when normalized by the torque $T_l = \rho \Omega^2 R_i^4 L/(\eta(1+\eta) Re)$ in the (single-phase) laminar regime,  satisfies the following relation (derived in the appendix):
\be\label{eq:rss}
\frac{T_b}{T_{l}} =   
1 + Re\ \frac{\eta}{1-\eta} \frac{1}{(\Omega R_i)^2} 
\int_{R_i}^{R_o}\!\!\!{\rm d}r
\left( \frac{\overline{u_r u_{\theta}}(r)}{r} + \left(\frac{r^2}{R_o^2}-1\right) \frac{\overline{ f_{b, \theta}}(r) }{2} \right),
\ee
where $\overline{u_r u_{\theta}}(r)$ is the azimuthal-radial component of the mean Reynolds shear stress and $\overline{ f_{b, \theta}}(r)$ the mean azimuthal component of bubble forcing.  Relation (\ref{eq:rss}) reveals the connection between the Reynolds shear stress, the bubble forcing, and the torque. We have tested that in the two-phase case the mean bubble forcing ${\bf f}_b$ is mainly dominated by the vertical component, see figure \ref{fig:fb}.
Therefore in the relation (\ref{eq:rss}) the contribution coming from  $\overline{ f_{b, \theta}}(r)$ is negligible compared to the Reynolds shear stress and the measure of $\overline{u_r u_{\theta}}(r)$ is a direct way to appreciate the variations in the mean flow features responsible for a torque change. 
As shown in Fig. \ref{fig:vprofiles} (bottom), this variation appears to be relevant in the whole bulk of the system and is approximately constant through it, both at large and small Reynolds numbers.
\begin{figure}
\begin{center}
\vspace{0.2cm}
\includegraphics[width=.48\textwidth,angle=0]{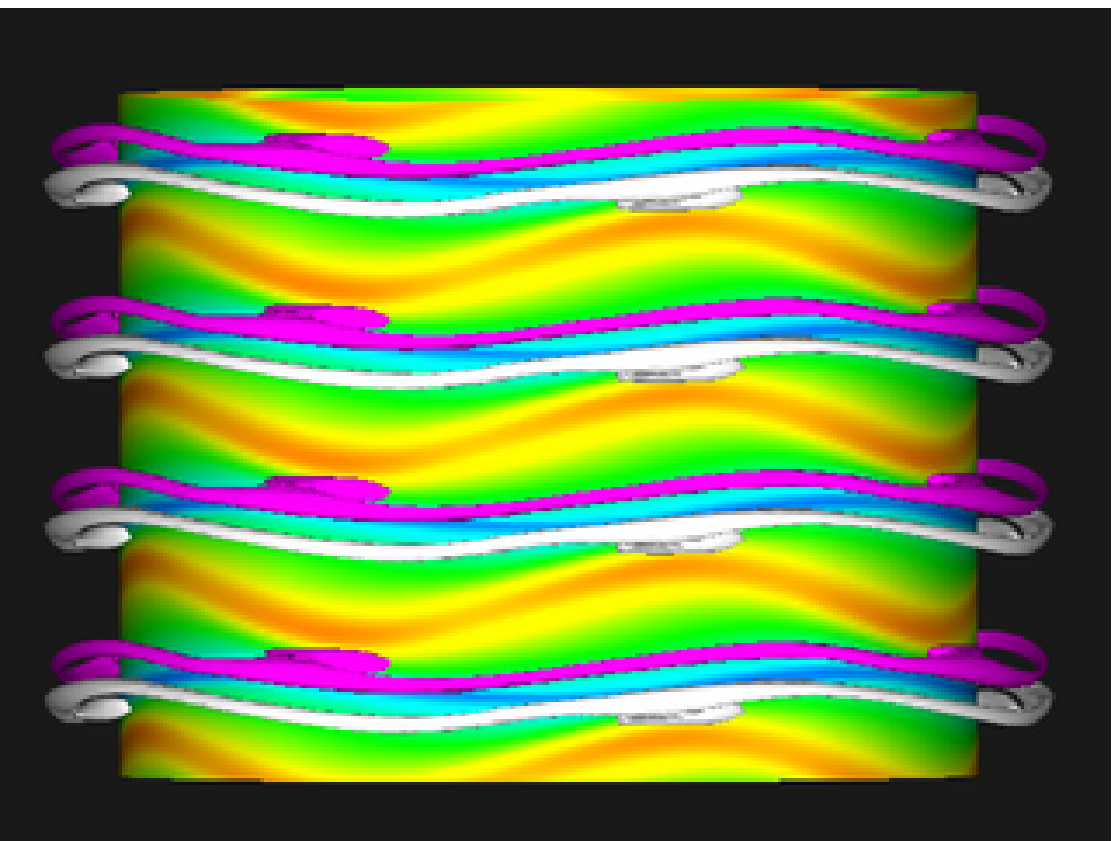}\hspace{0.1cm}
\includegraphics[width=.48\textwidth,angle=0]{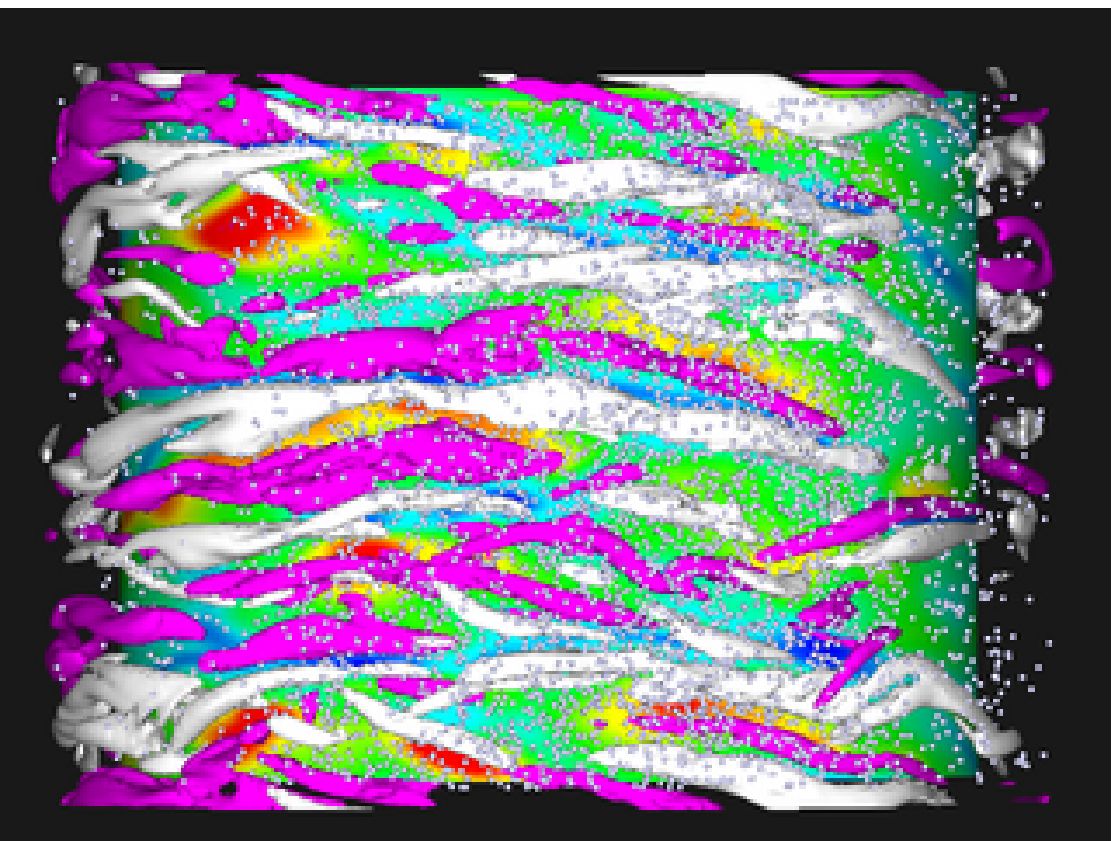}\\ \vspace{0.15cm}
\includegraphics[width=.48\textwidth,angle=0]{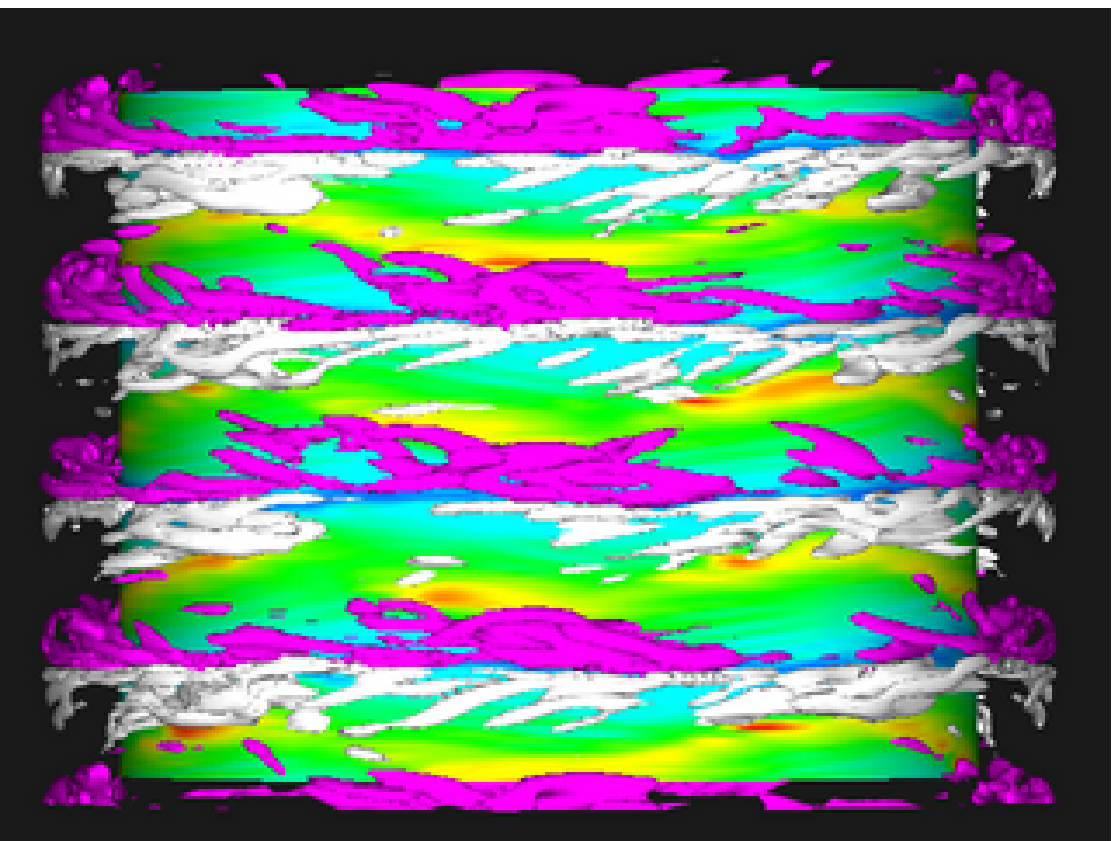}\hspace{0.1cm}
\includegraphics[width=.48\textwidth,angle=0]{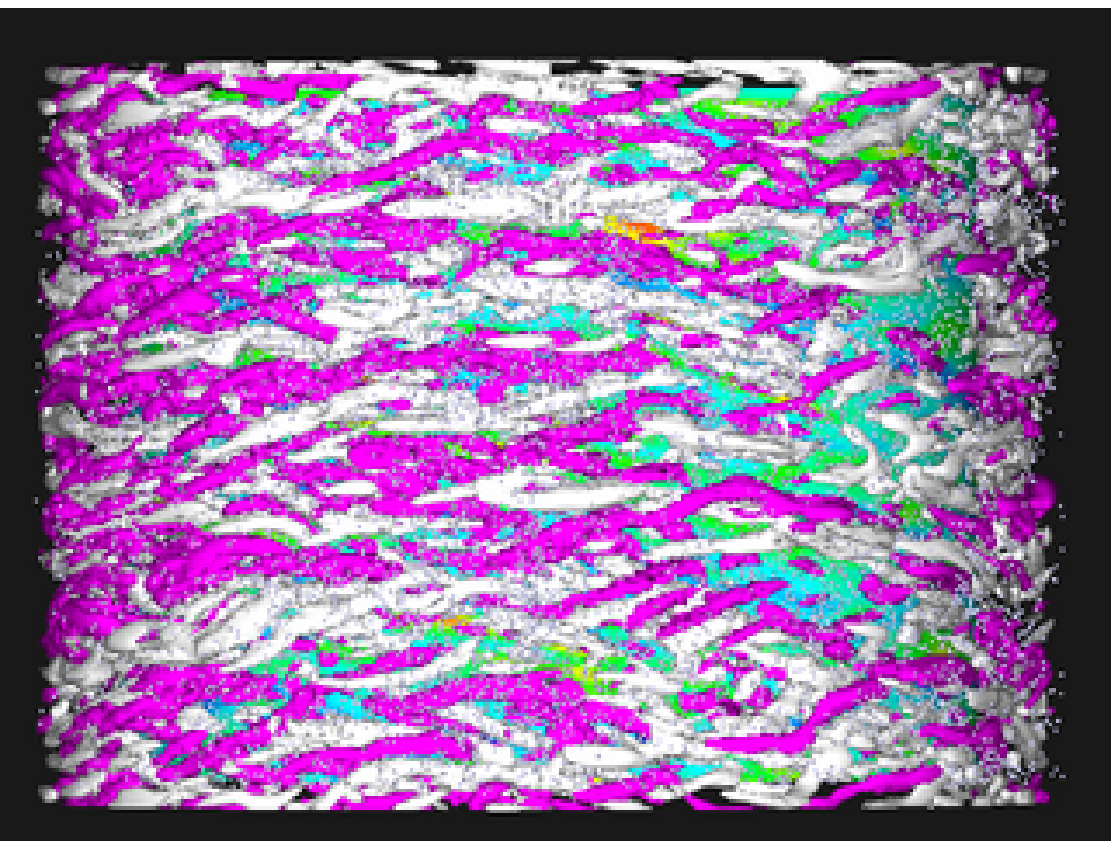}
\end{center}
\flushright \includegraphics[scale=.75,angle=0]{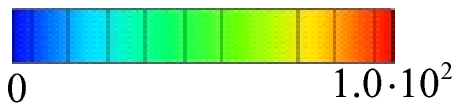}
\begin{center}
\caption{Flow visualization at $Re=900$ (top) and $Re=2000$ (bottom): Single-phase flow (left) and two-phase flow (right). 
The Taylor vortex structures are identified by iso-surfaces defined by the relation $(\partial_i u_j)(\partial_j u_i)=-0.4$, and uniformly colored (white and  violet) accordingly to the sign of $\omega_{\theta}$ on the surfaces. 
The wall shear stress, measured in $\rho (R_i \Omega)^2$ units, is reproduced in color on the inner cylinder surface.}\label{fig:vis}
\end{center}
\end{figure}

We also have previously observed that in the relative low Reynolds range under consideration, coherent flow structures in the form of Taylor vortices are persistent in the flow. To reveal the effect of the interaction of the dispersed bubbly phase with the coherent flow structures, we present a visualization of the flow. Figure \ref{fig:vis} shows snapshots from two simulations respectively at $Re=900$ and $Re=2000$, both in the single-phase and in two-phase case. 
Vortical structures are identified by iso-surfaces of the second invariant of the velocity gradient tensor and are shown in a uniform color accordingly to the sign of the azimuthal component of the local vorticity field, $\omega_{\theta}$  (\cite{hunt88}, \cite{tanahashi01}). The intensity of the inner-shear stress, $\tau_{w i}$, is also reproduced in colors in the figure. 
Vortical Taylor structures, that come in counter-rotating pairs, are well captured by this visualization method. In both cases it is evident how the addiction of microbubbles efficiently breaks the self organized order established by the system in the single-phase condition.

\subsection{Effect of bubbles on Taylor vortices}
We now have a more detail look at the Taylor vortices and how they are affected
by the bubble injection.
We note 
that in a periodic TC flow Taylor vortex rolls may appear only in even numbers. 
A Fourier spectral analysis of the azimuthal vorticity component on the mid-cylindrical surface of the gap, 
$\omega_{\theta}(r=R_i+d/2,z)$, reveals that the numerically simulated single-phase TC flow displays respectively 8 Taylor vortices for Reynolds values $Re \leq 2000$ and 6 above this threshold ($Re > 2000$).
The two-phase flow on the other hand shows a different behavior at small Reynolds number, in particular for $Re \leq 900$ we observe only 6  Taylor vortices, while the same number of rolls as the single-phase is realized for $Re>900$. This is qualitatively consistent with the experimental observation made by \cite{murai05}, as they observe an elongation of Taylor vortex spacing (i.e. the center-to-center distance) in the two-phase case as compared to the single-phase case. Such elongation decreases as $Re$ is increased. The single-phase/two-phase elongation ratio in the low-Reynolds number regime $Re\leq 900$ is roughly $30\%$ both in the simulations and in the experiment.
However, we cannot rule-out that the even-parity constrain imposed by
the periodic boundary condition in the DNS may introduce some kind of artifact on the overall stability of Taylor vortex couples.
Also, we note that the experimental measurement, see \cite{murai05}
figure 13, has been performed with larger bubbles ($a=1{\rm mm}$), as compared to the 
TRR measurements discussed in that paper ($a=0.3{\rm mm}$).
The bubbles employed to quantify the vortex elongation in the experiment turn out to correspond roughly to  $Re_b=400$ and $We=90$ and they are normally in a \textit{wobbling} state (\cite{grace73}). Such conditions lie outside the range of applicability of the Eulerian-Lagrangian model and they prevent from attempting a one-to-one comparison with the DNS.

We now focus on the azimuthal wave-length dependence in the DNS flow. We note that any signature of a dominant wave-length present in the single-phase case
is removed from the spectra of $\omega_{\theta}(r=R_i+d/2,z)$ in the two-phase flow case.  A transition from a clear azimuthal wave-length, that characterize the wavy-vortices and it is here of wave number 4, 
to a continuous and noisy-dominated spectra, is observed.

Again from this analysis we realize that the bubble phase seems to strongly break the coherence of the self-organized vortical structures, as observed in the previous flow visualizations (figure \ref{fig:vis}). 
Therefore, differently from the \cite{murai05} interpretation given for the experiment, we associate the observed two-phase flow drag reduction in the DNS to the breaking and consequent re-arrangement of coherent flow structures rather than to its stretching or displacement. This  view will be further supported by the discussion in the next subsection.

\subsection{Explanation for TRR decrease at increasing $Re$} \label{sec:decrease}
We now want to address the second of our questions, namely, why the drag reduction becomes less and less when  increasing the inner cylinder rotational speed.
Therefore we look at the global energy balance relation for the torque in the two-phase flow (see the appendix),
\bea\label{eq:Tb}
T_b &=&   \frac{\pi \rho (R_o^2 - R_i^2) L}{\Omega}\left( \langle \varepsilon_b \rangle  -  \langle {\bf f}_b \cdot  {\bf u} \rangle \right)\\
&=& \frac{T}{\langle \varepsilon \rangle} \left( \langle \varepsilon_b \rangle  - \langle {\bf f}_b \cdot  {\bf u} \rangle \right).\label{eq:Tb2}
\eea
The second equation (\ref{eq:Tb2}) follows from relation (\ref{eq:eps}).
\begin{figure}
\begin{center}
\includegraphics[width=.4\textwidth,angle=-90]{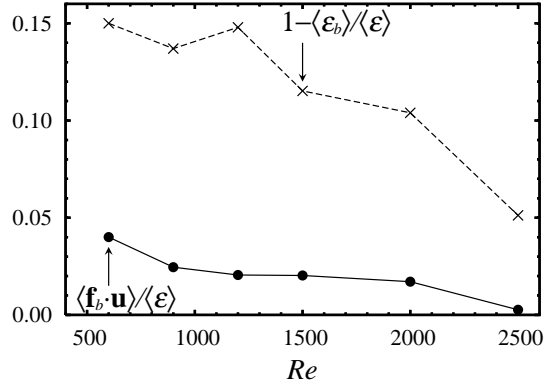}
\caption{The correlation coefficient $\langle {\bf f}_b \cdot  {\bf u} \rangle/ \langle \varepsilon \rangle$  and the relative variation of the mean energy dissipation rate of the two-phase flow as compared to the single-phase, 
$1 - \langle \varepsilon_b\rangle/\langle \varepsilon \rangle$ as function of $Re$. 
The sum of the two terms is the total torque reduction ratio (TRR).}\label{fig:corr}
\end{center}
\end{figure}
\begin{figure}
\begin{center}
\includegraphics[width=.4\textwidth,angle=-90]{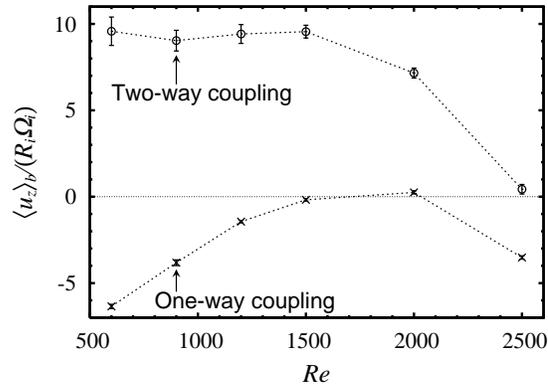}
\caption{
Mean vertical fluid velocity at the bubble positions, in dimensionless units $\langle u_z \rangle_b / (\Omega R_i)$ in a one-way and two-way run at variable $Re$. Bubbles stay preferentially in down-flow regions in the former case while is just the opposite for the latter.}\label{fig:corr1w}
\end{center}
\end{figure}
${\rm TRR}=1 - T/T_b$ is thus composed of the sum of the two terms 
\begin{equation}
{\rm TRR} = \left( 1-  {\langle \varepsilon_b \rangle \over \langle \varepsilon \rangle}
\right)   +
{\langle {\bf f}_b \cdot  {\bf u} \rangle \over \langle \varepsilon \rangle} ,
\label{sumof2terms}
\end{equation} 
which are here both of importance.
While the former is associated to changes of the dissipative structures in the system, the latter is associated to the extra energy input in the system due to the bubble momentum transfer. 
Both these terms are positive (i.e., contribute to the drag reduction). 
Furthermore, the $1-  \langle \varepsilon_b \rangle/\langle \varepsilon \rangle$ term is dominant at all $Re$, see figure \ref{fig:corr}. This means that the bubbly phase, that breaks vortical dissipation regions, modifies the energy flux directed from the large-scale of the external forcing (due to the rotating cylinder)  to the dissipative scale.
 
Both contributions $1-  \langle \varepsilon_b \rangle/\langle \varepsilon \rangle$ and $\langle {\bf f}_b \cdot  {\bf u} \rangle/\langle \varepsilon \rangle$ to TRR are vanishing at increasing $Re$. How to explain this dependence?
We note that $\langle {\bf f}_b \cdot  {\bf u} \rangle$, the correlation term of bubble forcing with velocity fluctuations, can be positive for two different reasons. First, the bubbles might distribute preferentially in regions where the velocity is up-flow as compared to the mean (and for clarity we note that the global mean velocity in the TC system is zero). This would produce a positive correlation with the term $- {\bf g}$  that is the dominant one in the bubble feed-back ${\bf f}_b$. The second possibility is that bubbles may induce a local up-flow fluctuation in the surrounding fluid that is larger as compared to the underlying flow fluctuation. This would produce always  a positive correlation with the gravity term $- {\bf g}$. The latter mechanism, where the bubble perturbation is overwhelming, is similar to what happens in a system where bubbles are rising in an otherwise quiescent fluid. Therefore, we will call this type of dynamics \textit{pseudo-turbulent}.

\begin{figure}
\begin{center}
\includegraphics[width=.6\textwidth,angle=-90]{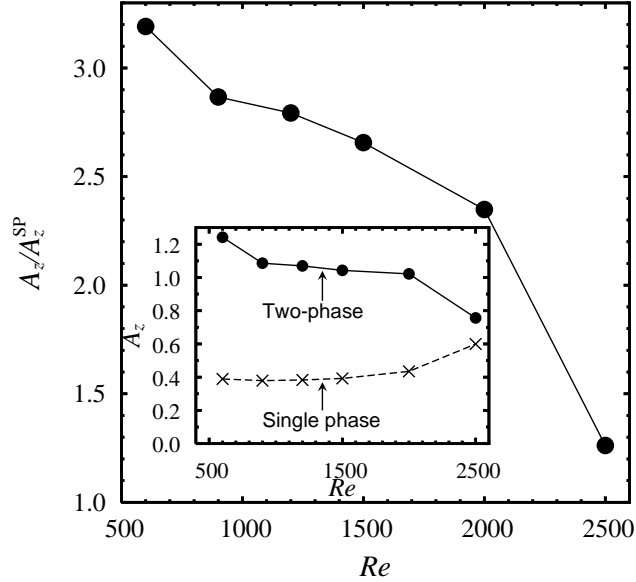}
\caption{The ratio  $A_z/A_z^{SP}$ versus $Re$, with $A_z = \langle u_z^{'2}\rangle/\langle u_{\theta}^{'2} + u_{r}^{'2}\rangle$
and ${\bf u'}=\langle ({\bf u} - \overline{{\bf u}}(r))^2 \rangle^{1/2}$ mean velocity fluctuation,
evaluated for the two-phase flow $A_z^{SP}$ represents the same quantity evaluated in the single-phase case.
In the inset, both  $A_z$ and $A_z^{SP}$ are shown separately as a function of $Re$.
}\label{fig:anis}
\end{center}
\end{figure}

In other words, given the assumption 
$\langle {\bf f}_b \cdot  {\bf u} \rangle \simeq - {\bf g}\cdot \langle {\bf u} \rangle_b$, where $\langle \ldots \rangle_b$ denotes the average on the bubble centroids, we would like to understand the behavior of $\langle {\bf u} \rangle_b$ in the different two-phase flow regimes. 
To distinguish between the two above mentioned possibilities, we make a numerical test  by switching-off the two-way coupling, i. e., ${\bf f}_b = 0$.  Also in this \textit{one-way} coupling case bubbles distribute inhomogeneously. 
In particular, we expect that the bubble distribution in the \textit{two-way} and \textit{one-way} coupling cases
shares similar properties if the amplitude of fluid fluctuations compared to the bubble feed-back is strong enough, while we expect them to have different properties if the pseudo-turbulence mechanism holds.  The evaluation of the quantity $\langle u_z \rangle_b$, in particular as a function of $Re$,
allows to discriminate between the two cases. 
In figure \ref{fig:corr1w} we show its behavior both in the one-way and two-way coupling case. 
In the one-way coupling case we observe preferentially bubble accumulation in down-flow regions. 
More precisely at low-$Re$, from flow and bubble visualization, we may observe that bubbles, while rising,  accumulate on stable regions in the flow. The phenomenology of the large-$Re$ regime is instead rather different, since the flow is time-dependent we do not observe bubble accumulation in stable positions but a rather homogeneous distribution that slightly favors downflow-regions. 
If we look now to the two-way coupled case we realize that bubbles strongly modify the flow by producing an upward flow in the position where they are.
Therefore this test supports the hypothesis of a pseudo-turbulent mechanism acting in the two-phase flow.
The pseudo-turbulent mechanism must become less effective at large $Re$ numbers because of the increased level of fluid fluctuations. 
This is confirmed by the measure of the intensity of the vertical anisotropic energy content in the system, figure \ref{fig:anis}. 
Vertical anisotropy in the two-phase flow, namely $A_z \equiv \langle u_z^{'2}\rangle/\langle u_{\theta}^{'2} + u_{r}^{'2}\rangle$ with ${\bf u'}=\langle ({\bf u} - \overline{{\bf u}}(r))^2 \rangle^{1/2}$ 
mean velocity fluctuation, when normalized by the anisotropy in the single-phase case, $A_z^{SP}$, monotonically decreases at increasing the Reynolds number.

To summarize, the mainly vertical extra-forcing exerted by the bubbles on the fluid is responsible for local and anisotropic modifications of the fluid fluctuations. No large-scale flow mechanism induced by bubbles seems to be present or relevant in this process. 
At increased external forcing (larger $Re$), the efficiency of the bubbles in perturbing the flow is lost.  
\subsection{Bubble distribution}\label{subsec:lift}
Inhomogeneous bubble distribution  in the TC system is due to the concurrence of several competing factors: (i) the centrifugal force, whose intensity towards the inner wall is proportional to $- u_{\theta}^2/r $; (ii) inertia that pushes the bubbles towards the local vortical flow structures (\cite{wang93}); (iii) the drift towards the walls, that is a lift effect due to the presence of a local mean shear (\cite{serizawa75}; and finally, (iv) the pseudo-turbulence flow fluctuations which may produce a weak side by side bubble-bubble attraction (\cite{zenit01}, \cite{bunner02}). All these effects are included explicitly, or via the bubble-flow interactions,  in the numerical model system.

\begin{figure}
\begin{center}
\includegraphics[width=.45\textwidth,angle=-90]{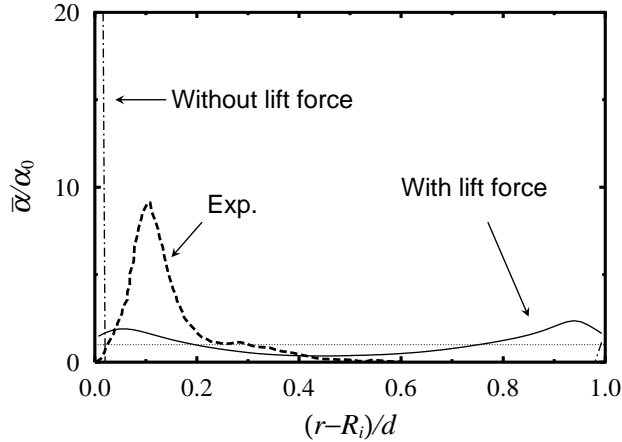}
\caption{Normalized radial mean void fraction at $Re=600$. Here a comparison of the numerical results with and without lift force are shown, together with the experimental profiles reported in \cite{murai05}. In the case without lift, nearly all the bubbles collapse into the first compartment of the histogram, giving a sharp bubble concentrations whose peak has been truncated for better readability.}\label{fig:lift1}
\end{center}
\end{figure}
For simplicity we limit our investigation to the radial dependence of mean bubble concentration profiles. 
The formations of vertical columnar or spiraling structures, similarly to the ones experimentally investigated by \cite{shiomi93}, are indeed observed in our numerics, but this will not be the focus of this section. 
Already from numerical results reported in figure \ref{fig:void} we have observed that at low $Re$ numbers bubbles accumulation at  both walls is observed, while at large $Re$ a weak tendency of accumulation in the core is noticeable.
Since in the low $Re$ condition  the drag behavior is dominated by pseudo-turbulence, we attribute near-wall accumulation to the bubble-wall interaction due to the lift in our numerical simulations. 
More explicitly, the local vertical mean flow produced by the bubble coupling is constrained by the no-slip conditions at the walls. This local up-flow profile, as in an up-flow vertical pipe (\cite{serizawa75}), induces a lateral (lift-driven) migration to the walls. Indeed this is evident if we switch off the lift force. Without lift all the bubbles collapse on the inner cylinder due to the centripetal force, see figure \ref{fig:lift1}.   
At large $Re$ the pseudo-turbulent effect is less efficient and the bubbles tend to accumulate in vortex core regions: inertia plays the dominant role. 
Figure  \ref{fig:lift1} also shows a comparison at $Re=600$ with the bubble concentration profile measured in the experiment. As noted previously, in the experimental profiles the local void fraction is always peaked near to the inner cylinder, in contrast to what is found in the simulations. 

What is the main reason for this discrepancy? 
First, the difference in statistically stationarity of the bubbly phase distribution. In the numerics the bubbles are injected statistically homogeneously into the fluid domain and a relatively long time, typically O(100) revolution times,  is required until statistically stationarity for the bubbles distribution is reached. In particular, after the bubbles release in the flow, we observe a transient behavior in their distribution. Bubbles first accumulate in separate columnar regions and then, when flow structures are further broken, they merge in larger domains.

Next, as we have already noted in the introduction, different bubble injection 
procedures are employed in the numerics as compared to the 
experiments. There, injection sites are localized at the bottom and near to the inner cylinder wall. From that positions bubbles spread to the upper section of the TC flow, where the bubble distributions are finally evaluated.
So, it could be that under the particular conditions of slow rotation of the inner cylinder (low $Re$) stationarity is not fully achieved in the experimental set-up, since the bubbles rise almost vertically. This seems to be supported by some experimental snapshots in \cite{murai05}, see figure 8 of that paper and the corresponding discussion.
Furthermore, bubble-bubble coalescence may sporadically occur in the experiment. Coalesced bubbles, which rise faster than the others, lead to the formation of clusters and may delay or prevent  the bubble distributions to become statistically stationary (Y. Murai 2007, personal communication).

\begin{figure}
\begin{center}
\includegraphics[width=.45\textwidth,angle=-90]{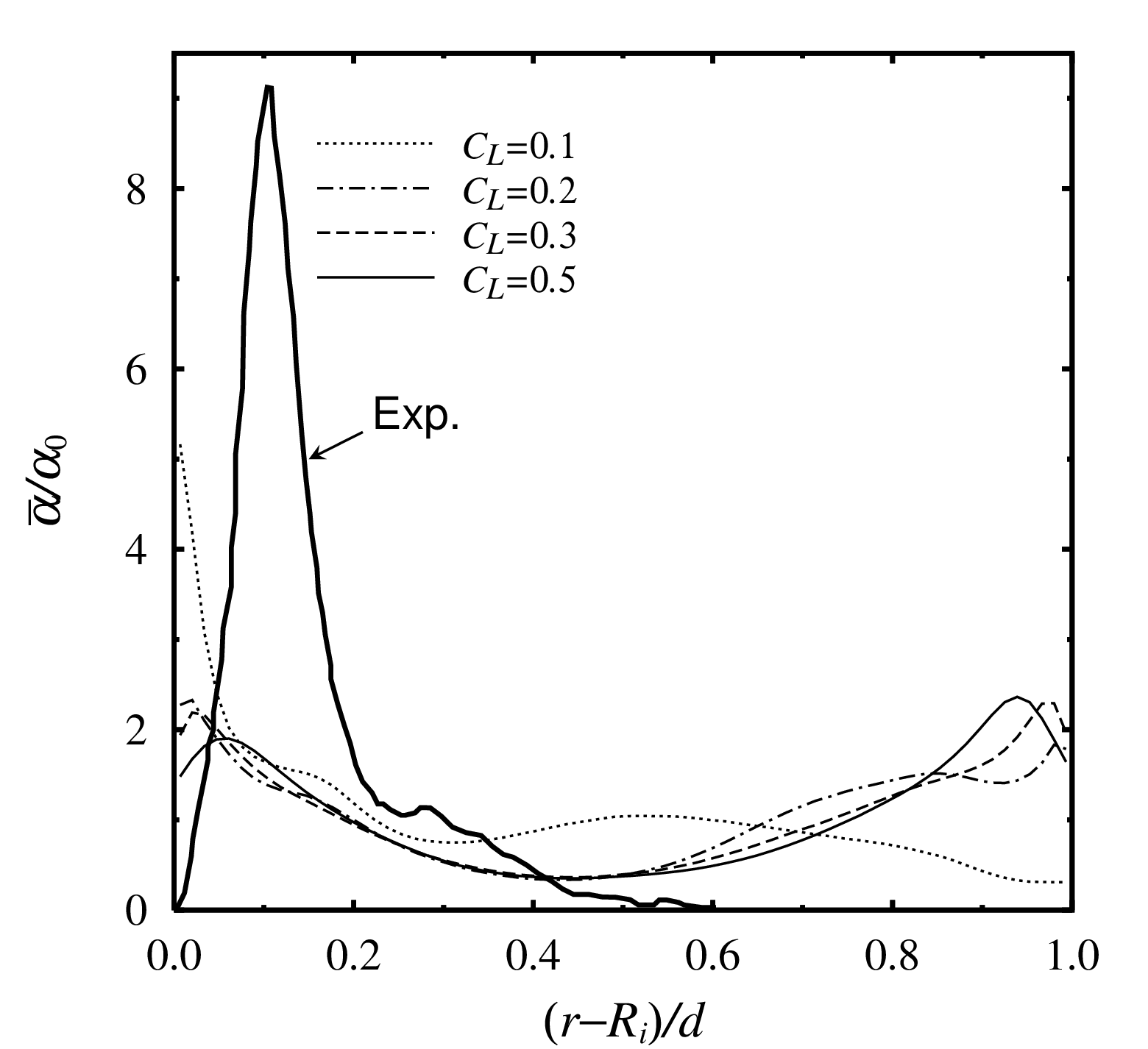}
\includegraphics[width=.45\textwidth,angle=-90]{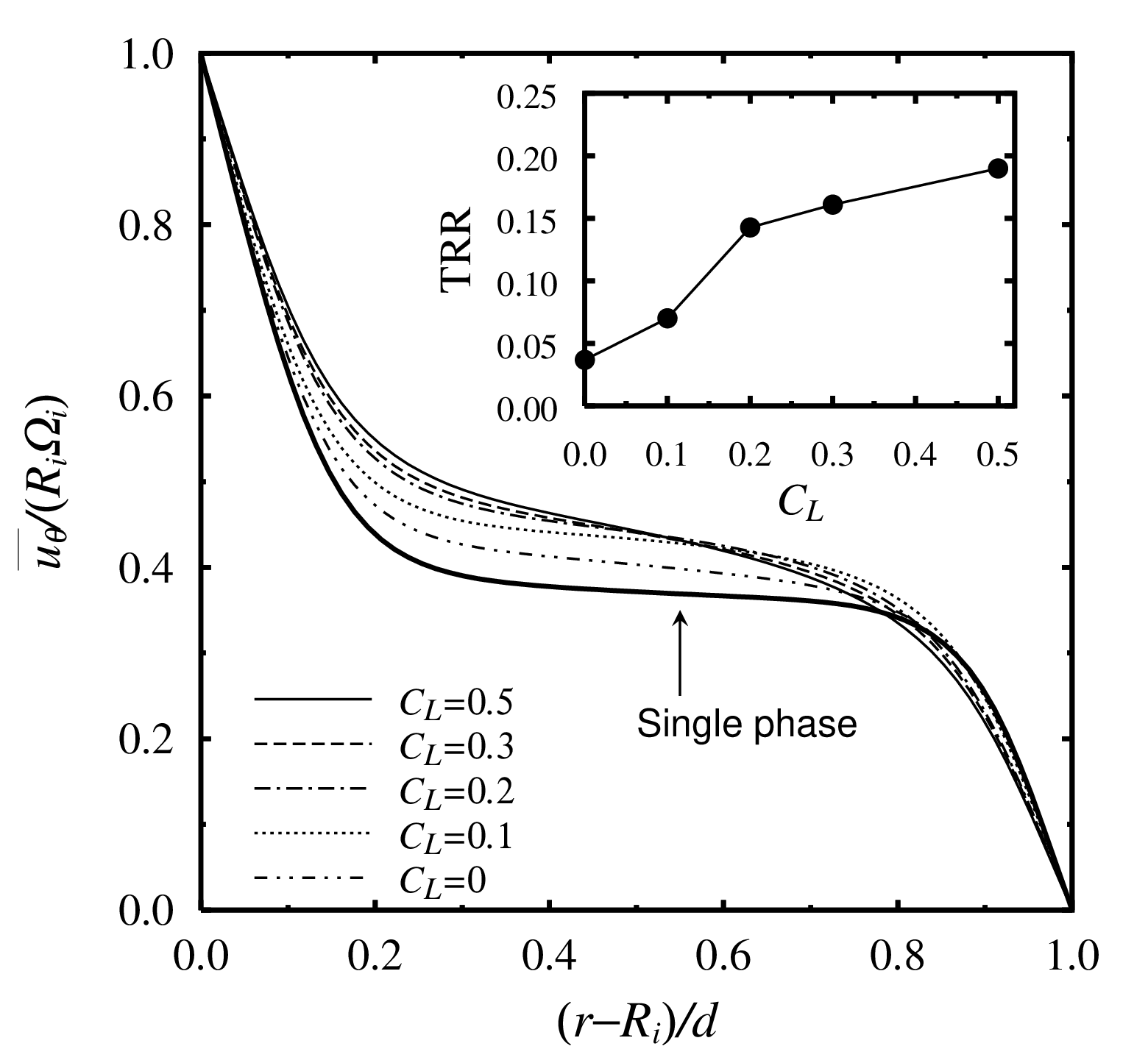}
\caption{Effect of the intensity of the lift force on the bubble mean radial distribution and on the drag reduction. 
The lift coefficient $C_L$ is varied in the interval $\left[0,0.5\right]$.
The left panel shows $\overline{\alpha}(r)/\alpha_{0}$ at changing $C_L$. On the right the corresponding shape of
the mean azimuthal velocity profile, $\overline{u_{\theta}}(r)$, is shown.
The inset reports the corresponding value of TRR versus $C_L$.}\label{fig:lift2}
\end{center}
\end{figure}

Second, on the numerical side, a possible source for the observed discrepancy may lie in the incompleteness of the model adopted for the lift force that may overestimate its relevance. 
It has been shown (\cite{magnaudet98}) that for non-deformable bubbles, whose Reynolds number ($Re_b$) is larger than one, and in a pure shear flow the lift coefficient can be considerably smaller than the asymptotic value, $C_L=1/2$. Similar findings for the lift coefficient have experimentally been obtained by \cite{nierop07}.  
Moreover, $C_L$ depends on the local shear and the local vorticity.
In order to understand how robust the shape of the mean bubble
concentration profile is when changing the lift coefficient $C_L$, we
have performed a test at the lowest Reynolds number ($Re=600$), where
the effect of bubble accumulation on the wall is more intense. $C_L$ has
been varied in the interval $\left[0,0.5\right]$ for a set of numerical
runs all starting from the same initial condition and extended until
statistically stationarity in time is reached. As shown in
Fig. \ref{fig:lift2}, it turns out that the bubbly mean concentration is
rather sensitive to $C_L$. For the cases with $C_L \leq 0.1$ it exhibits
a single wall-peak near the inner wall. This also implies a  dependence
on the bubble feed-back on the fluid. As reported in figure 
\ref{fig:lift2} (inset), reducing the lift coefficient progressively from $C_L=0.5$ to zero corresponds to a reduction in TRR. 
In particular, the accumulation of a small volume concentration of
bubbles (here $\alpha_0 = 0.125 \%$)  on the surface of the inner
cylinder, as for the cases $C_L=0$ and $C_L=0.1$, would have almost no
effect on the torque (or drag) modulation. 

\section{Conclusions}\label{sec:conclusions} 
We have numerically investigated the effect of microbubbles on a Taylor-Couette (TC) 
flow in the wavy vortex flow regime and focused on a comparison with a specific experiment where variations in the drag have been observed. 
We demonstrated that this phenomenon is reproduced by an Eulerian-Lagrangian model with point-force coupling. 
The main effect of microbubbles is to create a local perturbation of the flow, mainly directed upwards along the vertical direction that is able to break the coherent, and mainly dissipative, vortical structures of the flow. 
This pseudo-turbulent mechanism is at the origin of the torque reduction in the TC flow regime considered here.
As a consequence at large Reynolds number ($Re$), 
where the typical fluid velocity fluctuations are larger and coherent structures are unsteady and less persistent, this mechanism loses its efficiency.
The lift force resulted to be particularly important in this process. 
The strength of the lift force is the main factor for the bubble 
mean concentration profiles. Furthermore, the drag reduction is particularly sensitive to it.
When suppressing the lift, almost no drag reduction is obtained.

The dynamics responsible for the drag reduction in the low $Re$ TC flow highlighted in this study is fundamentally different from the one that has been observed in previous highly turbulent experimental investigations by \cite{vandenberg05,vandenberg07}. 
From our numerics 
it can be extrapolated that small non-deformable bubbles (zero Weber number ($We$)),
whose size is always below the smallest typical scale of the flow will be of less and less importance for drag reduction as the external forcing (and hence $Re$) is increased.
This is also consistent with the theory developed by \cite{lvov05}, where for small bubble in a channel flow the variation of the drag expected is only of the order of the void fraction.
This suggests that at larger $Re$, when  $We > 1$ (as in \cite{vandenberg05,vandenberg07}), other physical mechanisms shall be effective, namely, 
bubble deformation (\cite{lu05}) or compression (\cite{lo06}) or splitting (\cite{meng98}).
Further joint numerical and experimental works are still needed to clarify this aspect.\\
\emph{Acknowledgment:} 
We acknowledge Dr. Y. Murai for giving us some essential informations concerning his experimental setup 
and measurements and for reading the manuscript.

\begin{appendix}
\section{}\label{appendix}
\noindent
In this appendix we present a derivation of the equations
(\ref{eq:rss}) and (\ref{eq:Tb}), based on the use of the reciprocal
theorem (\cite{happelbrenner73}).
Let us introduce two incompressible velocity fields ${\bf u}$ and $\tilde{\bf u}$, both satisfying 
the same set of boundary conditions.
We identify ${\bf u}$ as the real flow in the TC system, and $\tilde{\bf u}$ as a particular flow field, that will be specified afterwards.
We focus here on the axially periodic TC system with only rotating inner cylinder. 
Therefore, the wall boundary conditions are the following:
\bea
{\bf u}=\tilde{\bf u}=R_i\Omega {\bf e}_\theta
\ \ \ &{\rm at}\ \ \ r=R_i, \nonumber \\
{\bf u}=\tilde{\bf u}=0
\ \ \ &{\rm at}\ \ \ r=R_o.
\label{eqap:bc01}
\eea
The instantaneous, i.e. time dependent,  torque $\mathcal{ T}$ acting on the inner cylinder is given by
\begin{equation}
\mathcal{T}\Omega=
R_i\Omega\oint_{r=R_i}\!\!\!\!\!\!{\rm d}^2{\bf x}\ 
(-{\bf e}_r\cdot {\bm \sigma}\cdot {\bf e}_\theta),
\label{eqap:trq01}
\end{equation}
where ${\bm \sigma}$ is the stress tensor associated to ${\bf u}$, and  ${\bf e}_r$ and ${\bf e}_\theta$
are unit vectors in the radial and azimuthal directions, respectively.
Using the boundary conditions (\ref{eqap:bc01})
for $\tilde{\bf u}$, we can rewrite (\ref{eqap:trq01})
in a volume integral form
\begin{equation}
\mathcal{T}\Omega=
\oint_{S}\!{\rm d}^2{\bf x}\ 
{\bf n}\cdot {\bm \sigma}\cdot \tilde{\bf u}=
\int_{V}\!{\rm d}^3{\bf x}\ 
(\nabla\cdot{\bm \sigma})\cdot\tilde{\bf u}
+
\int_{V}\!{\rm d}^3{\bf x}\ 
{\bm \sigma}:\nabla\tilde{\bf u}.
\label{eqap:trq02}
\end{equation}
Here $S$ is the area of the boundaries, 
$V$ is the fluid volume bounded by $S$, 
and the double-dot represents a product of two dyadics 
(see e.g. \cite{happelbrenner73}, Sec.2-1).
Substituting the momentum equation
$$
\nabla \cdot {\bm \sigma}=
\rho\left(
\partial_t{\bf u}+({\bf u}\cdot\nabla){\bf u}
-{\bf f}_b
\right),
$$
into (\ref{eqap:trq02}), we obtain
\begin{equation}
\mathcal{T}\Omega=
\rho\int_{V}\!{\rm d}^3{\bf x}
\left[
(\partial_t{\bf u})\cdot\tilde{\bf u}
-({\bf u}{\bf u}):\tilde{\bf S}
-{\bf f}_b\cdot\tilde{\bf u}
\right]
+\rho\oint_S\!{\rm d}^2{\bf x}\ 
\underbrace{({\bf n}\cdot{\bf u})}_{=0}
({\bf u}\cdot\tilde{\bf u})
+\int_{V}\!{\rm d}^3{\bf x}\ 
{\bm \sigma}:\nabla\tilde{\bf u},
\label{eqap:trq03}
\end{equation}
where ${\bf S}(=\{\nabla{\bf u}+(\nabla{\bf u})^T\}/2)$ 
denotes a strain tensor.
Similarly to the derivation of (\ref{eqap:trq02}), 
one can obtain
the inner torque $\tilde{\mathcal{T}}$ in the field 
$(\tilde{\bf u}, \tilde{\bm \sigma})$ 
\begin{equation}
\tilde{\mathcal{T}}\Omega=
\int_{V}\!{\rm d}^3{\bf x}\ 
(\nabla\cdot\tilde{\bm \sigma})\cdot{\bf u}
+
\int_{V}\!{\rm d}^3{\bf x}\ 
\tilde{\bm \sigma}:\nabla{\bf u}.
\label{eqap:trq04}
\end{equation}
We now apply the Lorentz reciprocal theorem 
$
{\bm \sigma}:\nabla\tilde{\bf u}
=
\tilde{\bm \sigma}:\nabla{\bf u}
$
(see e.g. \cite{happelbrenner73}  Sec. 3-5)
to (\ref{eqap:trq03}) and (\ref{eqap:trq04}), 
and rewrite the difference of the two torques 
in the volume integral form,
\begin{equation}
(\mathcal{T}-\tilde{\mathcal{T}})\Omega=
\rho\int_{V}\!{\rm d}^3{\bf x}
\left[
(\partial_t{\bf u})\cdot\tilde{\bf u}
-({\bf u}{\bf u}):\tilde{\bf S}
-{\bf f}_b\cdot\tilde{\bf u}
\right]
-
\int_{V}\!{\rm d}^3{\bf x}\ 
(\nabla\cdot\tilde{\bm \sigma})\cdot{\bf u}.
\label{eqap:trq05}
\end{equation}
From equation (\ref{eqap:trq03}) and (\ref{eqap:trq05}) the relevant relations (\ref{eq:rss}) and (\ref{eq:Tb}) can readily be obtained by proper assumptions on the field $\tilde{\bf u}$.
We consider two cases: (i) $\tilde{\bf u}$ is the real flow field ${\bf u}$, (ii) $\tilde{\bf u}$ is the steady flow field solution of the Stokes equation for the TC system.

\vspace{1em}

\noindent
{Case (i)}: 
If we choose the field $(\tilde{\bf u},\tilde{\bm \sigma})$ 
to be $({\bf u}, {\bm \sigma})$, 
the product $({\bm \sigma}:\nabla\tilde{\bf u})$
is equal to the energy dissipation rate $\varepsilon(=2\nu{\bf S}:{\bf S})$
and (\ref{eqap:trq03}) becomes
\begin{equation}
\mathcal{T}\Omega=
\frac{{\rm d}}{{\rm d}t}
\int_{V}\!{\rm d}^3{\bf x}\ 
\frac{\rho{\bf u}\cdot{\bf u}}{2}
+
\oint_{S}\!{\rm d}^2{\bf x}\ 
\underbrace{({\bf n}\cdot {\bf u})}_{=0}
\frac{\rho{\bf u}\cdot{\bf u}}{2}
+
\rho\int_{V}\!{\rm d}^3{\bf x}\ 
\left(
\varepsilon
-{\bf f}_b\cdot{\bf u}
\right),
\label{eqap:trq06}
\end{equation}
which corresponds to
the kinetic energy transport budget.
In a statistically steady state ($\langle \frac{ {\rm d} }{ {\rm d}t } \ldots \rangle=0$),
we obtain
\begin{equation}
\frac{T_b \Omega}{\pi\rho(R_o^2-R_i^2)L}
=\left\langle \varepsilon\right\rangle
-\left\langle{\bf f}_b\cdot{\bf u}\right\rangle,
\label{eqap:trq07}
\end{equation}
which accounts for the contribution of the energy dissipation rate and bubble forcing to the time averaged torque $T_b=\langle \mathcal{T} \rangle$.

\vspace{1em}

\noindent
{Case (ii)}: 
We now assume that the field $(\tilde{\bf u},\tilde{\bm \sigma})$ 
obeys the Stokes equation 
\begin{equation}
\nabla\cdot\tilde{\bm \sigma}=0.
\label{eqap:stokes01}
\end{equation}
Imposing the boundary conditions (\ref{eqap:bc01}), 
we write the solution of circular Couette flow (see \cite{chandra61}, Chapter VII)
\begin{equation}
\tilde{\bf u}=
{\bf e}_\theta
\left(
\frac{-R_ir+R_o^2R_ir^{-1}}{R_o^2-R_i^2}
\right)(R_i\Omega),
\quad \tilde{\bf S}=
-({\bf e}_r{\bf e}_\theta+{\bf e}_\theta{\bf e}_r)
\frac{R_o^2R_ir^{-2}}{R_o^2-R_i^2}
(R_i\Omega).
\label{eqap:stokes02}
\end{equation}
Substituting (\ref{eqap:stokes01}) and (\ref{eqap:stokes02})
into (\ref{eqap:trq05}), and assuming a statistically state,
we determine the mean torque
\begin{equation}
T_b
=
T_l
\left(
1+
\frac{Re(R_o^2-R_i^2)\eta}{2(R_i\Omega)^2(1-\eta)}
\left\langle
\frac{u_r u_\theta}{r^2}
-\frac{(R_o^2-r^2)f_{b,\theta}}{2R_o^2r}
\right\rangle
\right),
\label{eq:fiktc01}
\end{equation}
where $T_l$ denotes the torque in the laminar flow,
\begin{equation}
T_l=
\frac{4\pi\rho (R_i\Omega)^2R_i^2L}{Re\ \eta(1+\eta)}.
\end{equation}
The first term inside the brackets on the right hand side of (\ref{eq:fiktc01})
accounts for the laminar flow contribution
and the second term 
for the modification from the circular Couette flow
due to the nonlinearity of the fluid motion
and the bubble's forcing.
Using the area averaged quantities
$\overline{u_r u_\theta}(r)$ and $\overline{f_{b,\theta}}(r)$,
we may finally write (\ref{eq:fiktc01}) in the radial integral form, previously used in this paper,
\begin{equation}
T_b
=
T_l
\left(
1
+\frac{Re\ \eta}{(R_i\Omega)^2(1-\eta)}
\int_{R_i}^{R_o}\!\!\!{\rm d}r\ 
\left(
\frac{\overline{u_r u_\theta}(r)}{r}
-\frac{(R_o^2-r^2)\overline{f_{b, \theta}}(r)}{2R_o^2}
\right)
\right).
\label{eq:fiktc02}
\end{equation}
We note that a general identity linking the effect of the Reynolds shear stress distribution on the skin friction
has been recently derived  by  \cite{fukagata02b}  for the case of a pressure driven channel and circular pipe flows.   
The Fukagata \textit{et al.} identity has been further extended to the problem of channel flow with mixed boundary conditions 
by using the reciprocal theorem in \cite{sbragaglia07}. 
Our derivation presented above represents an extension of that work to the TC system.
\end{appendix}

\bibliographystyle{jfm}

\begin{thebibliography}{46}
\expandafter\ifx\csname natexlab\endcsname\relax\def\natexlab#1{#1}\fi

\bibitem[Andereck {\em et~al.\/}(1986)Andereck, Liu \& Swinney]{andereck86}
{\sc Andereck, C.~D., Liu, S.~S. \& Swinney, H.~L.} 1986 {Flow regimes in a
  circular Couette system with independently rotating cylinders}. {\em J. Fluid
  Mech.\/} {\bf 164}, 155--183.

\bibitem[Auton(1987)]{auton87}
{\sc Auton, T.~R.} 1987 {The lift force on a spherical body in a rotational
  flow}. {\em J. Fluid Mech.\/} {\bf 183}, 199--218.

\bibitem[Auton {\em et~al.\/}(1988)Auton, Hunt \& Prud'homme]{auton88}
{\sc Auton, T.~R., Hunt, J. C.~R. \& Prud'homme, M.} 1988 {The force exerted on
  a body in inviscid unsteady non-uniform rotational flow}. {\em J. Fluid
  Mech.\/} {\bf 197}, 241--257.

\bibitem[van~den Berg {\em et~al.\/}(2007)van~den Berg, van Gils, Lathrop \&
  Lohse]{vandenberg07}
{\sc van~den Berg, T.~H., van Gils, D. P.~M., Lathrop, D.~P. \& Lohse, D.} 2007
  {Bubbly turbulent drag reduction is a boundary layer effect}. {\em Phys. Rev.
  Lett.\/} {\bf 98}, 084501.

\bibitem[van~den Berg {\em et~al.\/}(2005)van~den Berg, Luther, Lathrop \&
  Lohse]{vandenberg05}
{\sc van~den Berg, T.~H., Luther, S., Lathrop, D.~P. \& Lohse, D.} 2005 {Drag
  reduction in bubbly Taylor-Couette turbulence}. {\em Phys. Rev. Lett.\/} {\bf
  94}, 044501.

\bibitem[van~den Berg {\em et~al.\/}(2006)van~den Berg, Luther \&
  Lohse]{ramon06b}
{\sc van~den Berg, T.~H., Luther, S. \& Lohse, D.} 2006 {Energy spectra in
  microbubbly turbulence}. {\em Phys. Fluids\/} {\bf 18}, 038103.

\bibitem[Bunner \& Tryggvason(2002)]{bunner02}
{\sc Bunner, B. \& Tryggvason, G.} 2002 {Dynamics of homogeneous bubbly flows
  Part 1. Rise velocity and microstructure of the bubbles}. {\em J. Fluid
  Mech.\/} {\bf 466}, 17--52.

\bibitem[Calzavarini {\em et~al.\/}(2006)Calzavarini, van~den Berg, Luther,
  Toschi \& Lohse]{calza06}
{\sc Calzavarini, E., van~den Berg, T.~H., Luther, S., Toschi, F. \& Lohse, D.}
  2006 {Microbubble clustering in a turbulent flow}. {\em
  http://lanl.arxiv.org/abs/physics/0607255\/} .

\bibitem[Chandrasekhar(1961)]{chandra61}
{\sc Chandrasekhar, S.} 1961 {\em {Hydrodynamic and Hydromagnetic
  Stability}\/}. Clarendon Press (Oxford).

\bibitem[Climent \& Magnaudet(1999)]{climent99}
{\sc Climent, E. \& Magnaudet, J.} 1999 {Large-scale simulations of
  bubble-induced convection in a liquid layer}. {\em Phys. Rev. Lett.\/} {\bf
  82}, 4827--4830.

\bibitem[Djeridi {\em et~al.\/}(2004)Djeridi, Gabillet \& Billard]{djeridi04}
{\sc Djeridi, H., Gabillet, C. \& Billard, J.~Y.} 2004 {Two-phase
  Couette--Taylor flow: Arrangement of the dispersed phase and effects on the
  flow structures}. {\em Phys. Fluids\/} {\bf 16}, 128--139.

\bibitem[Dominguez-Lerma {\em et~al.\/}(1984)Dominguez-Lerma, Ahlers \&
  Cannell]{dom84}
{\sc Dominguez-Lerma, M.~A., Ahlers, G. \& Cannell, D.} 1984 {Marginal
  stability curve and linear growth rate for rotating Couette--Taylor flow and
  Rayleigh--B{\'e}nard convection}. {\em Phys. Fluids\/} {\bf 27}, 856--860.

\bibitem[Dutcher \& Muller(2007)]{dutch07}
{\sc Dutcher, C.~S. \& Muller, S.~J.} 2007 {Explicit analytic formulas for
  Newtonian Taylor-Couette primary instabilities}. {\em Phys. Rev. E\/} {\bf
  75}, 047301.

\bibitem[Eckhardt {\em et~al.\/}(2000)Eckhardt, Grossmann \& Lohse]{eckhardt00}
{\sc Eckhardt, B., Grossmann, S. \& Lohse, D.} 2000 {Scaling of global momentum
  transport in Taylor-Couette and pipe flow}. {\em Eur. Phys. J. B\/} {\bf 18},
  541--544.

\bibitem[Eckhardt {\em et~al.\/}(2007)Eckhardt, Grossmann \& Lohse]{gross07}
{\sc Eckhardt, B., Grossmann, S. \& Lohse, D.} 2007 {Torque scaling in
  turbulent Taylor-Couette flow between independently rotating cylinders}. {\em
  J. Fluid Mech\/} {\bf 581}, 221--250.

\bibitem[Esser \& Grossmann(1996)]{esser96}
{\sc Esser, A. \& Grossmann, S.} 1996 {Analytic expression for Taylor-Couette
  stability boundary}. {\em Phys. Fluids\/} {\bf 8}, 1814--1819.

\bibitem[Ferrante \& Elghobashi(2004)]{fer04}
{\sc Ferrante, A. \& Elghobashi, S.} 2004 On the physical mechanisms of drag
  reduction in a spatially-developing turbulent boundary layer laden with
  microbubbles. {\em J. Fluid Mech.\/} {\bf 503}, 345--355.

\bibitem[Fukagata {\em et~al.\/}(2002)Fukagata, Iwamoto \& Kasagi]{fukagata02b}
{\sc Fukagata, K., Iwamoto, K. \& Kasagi, N.} 2002 {Contribution of Reynolds
  stress distribution to the skin friction in wall-bounded flows}. {\em Phys.
  Fluids\/} {\bf 14}, L73--L76.

\bibitem[Fukagata \& Kasagi(2002)]{fukagata02}
{\sc Fukagata, K. \& Kasagi, N.} 2002 {Highly energy-conservative finite
  difference method for the cylindrical coordinate system}. {\em J. Comp.
  Phys.\/} {\bf 181}, 478--498.

\bibitem[Grace(1973)]{grace73}
{\sc Grace, J.} 1973 Shapes and velocities of bubbles rising in infinite
  liquids. {\em Trans. Instn. Chem. Eng.\/} {\bf 51}, 116--120.

\bibitem[Happel \& Brenner(1973)]{happelbrenner73}
{\sc Happel, J. \& Brenner, H.} 1973 {\em {Low Reynolds Number
  Hydrodynamics}\/}. (Second edition) Martinus Nijhoff Publishers.

\bibitem[Hunt {\em et~al.\/}(1988)Hunt, Wray \& Moin]{hunt88}
{\sc Hunt, J. C.~R., Wray, A.~A. \& Moin, P.} 1988 {Eddies, stream and
  convergence zones in turbulent flows., Report CTR-S88}. {\em Center for
  Turbulence Research, NASA Ames Research Center and Stanford University,
  California, USA\/} .

\bibitem[Lathrop {\em et~al.\/}(1992)Lathrop, Fineberg \& Swinney]{lathrop92}
{\sc Lathrop, D.~P., Fineberg, J. \& Swinney, H.~L.} 1992 {Transition to
  shear-driven turbulence in Couette-Taylor flow}. {\em Phys. Rev. A\/} {\bf
  46}, 6390--6405.

\bibitem[Lim \& Tan(2004)]{lim04}
{\sc Lim, T.~T. \& Tan, K.~S.} 2004 {A note on power-law scaling in a
  Taylor-Couette flow}. {\em Phys. Fluids\/} {\bf 16}, 140--144.

\bibitem[Lo {\em et~al.\/}(2006)Lo, L'vov \& Procaccia]{lo06}
{\sc Lo, T.~S., L'vov, V.~S. \& Procaccia, I.} 2006 {Drag reduction by
  compressible bubbles}. {\em Phys. Rev. E\/} {\bf 73}, 036308.

\bibitem[Lu {\em et~al.\/}(2005)Lu, Fern{\'a}ndez \& Tryggvason]{lu05}
{\sc Lu, J., Fern{\'a}ndez, A. \& Tryggvason, G.} 2005 {The effect of bubbles
  on the wall drag in a turbulent channel flow}. {\em Phys. Fluids\/} {\bf 17},
  095102.

\bibitem[L'vov {\em et~al.\/}(2005)L'vov, Pomyalov, Procaccia \&
  Tiberkevich]{lvov05}
{\sc L'vov, V.~S., Pomyalov, A., Procaccia, I. \& Tiberkevich, V.} 2005 {Drag
  reduction by microbubbles in turbulent flows: The limit of minute bubbles}.
  {\em Phys. Rev. Lett.\/} {\bf 94}, 174502.

\bibitem[Madavan {\em et~al.\/}(1984)Madavan, Deutsch \& Merkle]{madavan84}
{\sc Madavan, N.~K., Deutsch, S. \& Merkle, C.~L.} 1984 {Reduction of turbulent
  skin friction by microbubbles}. {\em Phys. Fluids\/} {\bf 27}, 356--363.

\bibitem[Magnaudet \& Legendre(1998)]{magnaudet98}
{\sc Magnaudet, J. \& Legendre, D.} 1998 {Some aspects of the lift force on a
  spherical bubble}. {\em App. Sci. Res.\/} {\bf 58}, 441--461.

\bibitem[Maxey \& Riley(1983)]{maxeyriley83}
{\sc Maxey, M.~R. \& Riley, J.~J.} 1983 {Equation of motion for a small rigid
  sphere in a nonuniform flow}. {\em Phys. Fluids\/} {\bf 26}, 883--889.

\bibitem[Mazzitelli {\em et~al.\/}(2003{\natexlab{{\em a\/}}})Mazzitelli, Lohse
  \& Toschi]{mazzitelli03}
{\sc Mazzitelli, I.~M., Lohse, D. \& Toschi, F.} 2003{\natexlab{{\em a\/}}} {On
  the relevance of the lift force in bubbly turbulence}. {\em J. Fluid Mech.\/}
  {\bf 488}, 283--313.

\bibitem[Mazzitelli {\em et~al.\/}(2003{\natexlab{{\em b\/}}})Mazzitelli, Lohse
  \& Toschi]{mazzitelli03b}
{\sc Mazzitelli, I.~M., Lohse, D. \& Toschi, F.} 2003{\natexlab{{\em b\/}}}
  {The effect of microbubbles on developed turbulence}. {\em Phys. Fluids\/}
  {\bf 15}, L5--L8.

\bibitem[Meng \& Uhlman(1998)]{meng98}
{\sc Meng, J. C.~S. \& Uhlman, J.~S.} 1998 {Microbubble formation and splitting
  in a turbulent boundary layer for turbulence reduction}. {\em Proceedings of
  the International Symposium on Seawater Drag Reduction, Newport, RI, July, US
  Office of Naval Research, Arlington, VA\/} pp. 341--355.

\bibitem[Murai {\em et~al.\/}(2005)Murai, Oiwa \& Takeda]{murai05}
{\sc Murai, Y., Oiwa, H. \& Takeda, Y.} 2005 {Bubble behavior in a vertical
  Taylor-Couette flow}. {\em J. Phys. Conf. Ser.\/} {\bf 14}, 143--156.

\bibitem[van Nierop {\em et~al.\/}(2007)van Nierop, Luther, Bluemink,
  Magnaudet, Prosperetti \& Lohse]{nierop07}
{\sc van Nierop, E.~A., Luther, S., Bluemink, J.~J., Magnaudet, J.,
  Prosperetti, A. \& Lohse, D.} 2007 {Drag and lift forces on bubbles in a
  rotating flow}. {\em J. Fluid Mech.\/} {\bf 571}, 439--454.

\bibitem[Oiwa(2005)]{oiwa05}
{\sc Oiwa, H.} 2005 {A study of mechanisms modifying shear stress in bubbly
  flow}. {\em Doctoral Thesis, Fukui University (Japan)\/} .

\bibitem[Sanders {\em et~al.\/}(2006)Sanders, Winkel, Dowling, Perlin \&
  Ceccio]{sanders06}
{\sc Sanders, W.~C., Winkel, E.~S., Dowling, D.~R., Perlin, M. \& Ceccio,
  S.~L.} 2006 {Bubble friction drag reduction in a high-Reynolds-number
  flat-plate turbulent boundary layer}. {\em J. Fluid Mech.\/} {\bf 552},
  353--380.

\bibitem[Sbragaglia \& Sugiyama(2007)]{sbragaglia07}
{\sc Sbragaglia, M. \& Sugiyama, K.} 2007 {Boundary induced nonlinearities at
  small Reynolds numbers}. {\em Physica D\/} {\bf 228}, 140--147.

\bibitem[Serizawa {\em et~al.\/}(1975)Serizawa, Kataoka \&
  Michiyoshi]{serizawa75}
{\sc Serizawa, A., Kataoka, I. \& Michiyoshi, I.} 1975 {Turbulence structure of
  air-water bubbly flow - II. Local properties}. {\em Int. J. Mult. Flow\/}
  {\bf 2}, 235--246.

\bibitem[Shiomi {\em et~al.\/}(1993)Shiomi, Kutsuma, Akagawa \&
  Ozawa]{shiomi93}
{\sc Shiomi, Y., Kutsuma, H., Akagawa, K. \& Ozawa, M.} 1993 {Two-phase flow in
  an annulus with a rotating inner cylinder (flow pattern in bubbly flow
  region)}. {\em Nucl. Eng. Des.\/} {\bf 141}, 27--34.

\bibitem[Tanahashi {\em et~al.\/}(2001)Tanahashi, Iwase \&
  Miyauchi]{tanahashi01}
{\sc Tanahashi, M., Iwase, S. \& Miyauchi, T.} 2001 {Appearance and alignment
  with strain rate of coherent fine scale eddies in turbulent mixing layer}.
  {\em J. Turb.\/} {\bf 2}, 1--17.

\bibitem[Taylor(1923)]{taylor23}
{\sc Taylor, G.~I.} 1923 {Stability of a viscous liquid contained between two
  rotating cylinders}. {\em Phil. Trans. Roy. Soc. A\/} {\bf 223}, 289--343.

\bibitem[Wang \& Maxey(1993)]{wang93}
{\sc Wang, L.~P. \& Maxey, M.~R.} 1993 {The motion of microbubbles in a forced
  isotropic and homogeneous turbulence}. {\em Appl. Sci. Res.\/} {\bf 51},
  291--296.

\bibitem[Wendt(1933)]{wendt33}
{\sc Wendt, F.} 1933 {Turbulente Str\"{o}mungen zwischen zwei rotierenden
  konaxialen Zylindern}. {\em Ingenieurs-Archiv\/} {\bf 4}, 577--595.

\bibitem[Xu {\em et~al.\/}(2002)Xu, Maxey \& Karniadakis]{xu02}
{\sc Xu, J., Maxey, M.~R. \& Karniadakis, G. E.~M.} 2002 {Numerical simulation
  of turbulent drag reduction using micro-bubbles}. {\em J. Fluid Mech.\/} {\bf
  468}, 271--281.

\bibitem[Zenit {\em et~al.\/}(2001)Zenit, Koch \& Sangani]{zenit01}
{\sc Zenit, R., Koch, D.~L. \& Sangani, A.~S.} 2001 {Measurements of the
  average properties of a suspension of bubbles rising in a vertical channel}.
  {\em J. Fluid Mech.\/} {\bf 429}, 307--342.

\end{thebibliography}

\end{document}